\documentclass{aastex631}
\usepackage{float}

\newcommand{\kR}{\mbox{$\kappa_{\rm R}$}}

\newcommand{\Msun}{\mbox{$M_{\odot}$}}

\newcommand{\Teff}{\mbox{$T_{\rm eff}$}}

\received{May 16, 2024}
\accepted{September 14, 2024}

\shorttitle{Opacities at  High Pressure}
\shortauthors{Marigo et al.}
\graphicspath{{./}{figures/}}
\usepackage{multirow}
\usepackage{booktabs}
\usepackage{threeparttable}
\usepackage{courier}

\begin{document}
\title{\AE SOPUS 2.1: Low-Temperature Opacities Extended to High Pressure}

\correspondingauthor{Diego Bossini}
\email{diego.bossini@unipd.it}

\author[0000-0002-9137-0773]{Paola Marigo}
\affiliation{Department of Physics and Astronomy G. Galilei,
University of Padova, Vicolo dell'Osservatorio 3, I-35122, Padova, Italy}
\author[0000-0002-3867-9966]{Francesco Addari}
\affiliation{Scuola Internazionale Superiore di Studi Avanzati, Via Bonomea, 265, I-34136, Trieste, Italy}
\author[0000-0002-9480-8400]{Diego Bossini}
\affiliation{Department of Physics and Astronomy G. Galilei, University of Padova, Vicolo dell'Osservatorio 3, I-35122, Padova, Italy}
\author[0000-0002-7922-8440]{Alessandro Bressan}
\affiliation{Scuola Internazionale Superiore di Studi Avanzati, Via Bonomea, 265, I-34136, Trieste, Italy}
\author[0000-0002-6213-6988]{Guglielmo Costa}
\affiliation{Univ Lyon, Univ Lyon1, Ens de Lyon, CNRS, Centre de Recherche Astrophysique de Lyon UMR5574, \\F-69230 Saint-Genis-Laval, France}
\affiliation{INAF-Osservatorio Astronomico di Padova, Vicolo dell’Osservatorio 5, I-35122 Padova, Italy}
\author[0000-0002-6301-3269]{L\'eo Girardi}
\affiliation{INAF-Osservatorio Astronomico di Padova, Vicolo dell’Osservatorio 5, I-35122 Padova, Italy}
\author[0000-0002-1429-2388]{Michele Trabucchi}
\affiliation{Department of Physics and Astronomy G. Galilei, University of Padova, Vicolo dell'Osservatorio 3, I-35122, Padova, Italy}
\affiliation{INAF-Osservatorio Astronomico di Padova, Vicolo dell’Osservatorio 5, I-35122 Padova, Italy}
\author[0000-0002-8691-4940]{Guglielmo Volpato}
\affiliation{Department of Physics and Astronomy G. Galilei, University of Padova, Vicolo dell'Osservatorio 3, I-35122, Padova, Italy}
\affiliation{INAF-Osservatorio Astronomico di Padova, Vicolo dell’Osservatorio 5, I-35122 Padova, Italy}



\begin{abstract}
We address the critical need for accurate Rosseland mean gas opacities in high-pressure environments, spanning temperatures from 100 K to 32000 K. Current opacity tables from Wichita State University and \texttt{\AE SOPUS~2.0} are limited to $\log(R) \le 1$, where $R=\rho\, T_6^{-3}$ in units of $\mathrm{g}\,\mathrm{cm}^{-3}(10^6\mathrm{K})^{-3}$. This is insufficient for modeling very low-mass stars, brown dwarfs, and planets with atmospheres exhibiting higher densities and pressures ($\log(R) > 1$). Leveraging extensive databases such as \texttt{ExoMol}, \texttt{ExoMolOP}, \texttt{MoLLIST}, and \texttt{HITEMP}, we focus on expanding the \texttt{\AE SOPUS} opacity calculations to cover a broad range of pressure and density conditions ($-8 \leq \log(R) \leq +6$). 
We incorporate the thermal Doppler mechanism and micro-turbulence velocity. Pressure broadening effects on molecular transitions, leading to Lorentzian or Voigt profiles, are explored in the context of atmospheric profiles for exoplanets, brown dwarfs, and low-mass stars. We also delve into the impact of electron degeneracy and non-ideal effects such as ionization potential depression under high-density conditions, emphasizing its notable influence on Rosseland mean opacities at temperatures exceeding $10,000$ K.
As a result, this study expands \texttt{\AE SOPUS} public web interface for customized gas chemical mixtures, promoting flexibility in opacity calculations based on specific research needs. Additionally, pre-computed opacity tables, inclusive of condensates, are provided. We present a preliminary application to evolutionary models for very low-mass stars.
\end{abstract}



\section{Introduction} \label{sec_intro}
Thanks to extensive databases such as \texttt{ExoMol} \citep{EXOMOL_2012MNRAS.425...21T, Tennyson_etal_16}, \texttt{ExoMolOP} \citep{Chubb_etal_21}, \texttt{MoLLIST} \citep{Bernath_etal_20} and \texttt{HITEMP} \citep{Rothman_etal_10}, we can now rely on the availability of a substantial amount of molecular line list data,  which is a significant asset for modeling the atmospheres of hot exoplanets, as well as cool stellar and sub-stellar atmospheres.
\begin{figure}[h]
    \centering    \includegraphics[width=0.5\textwidth]{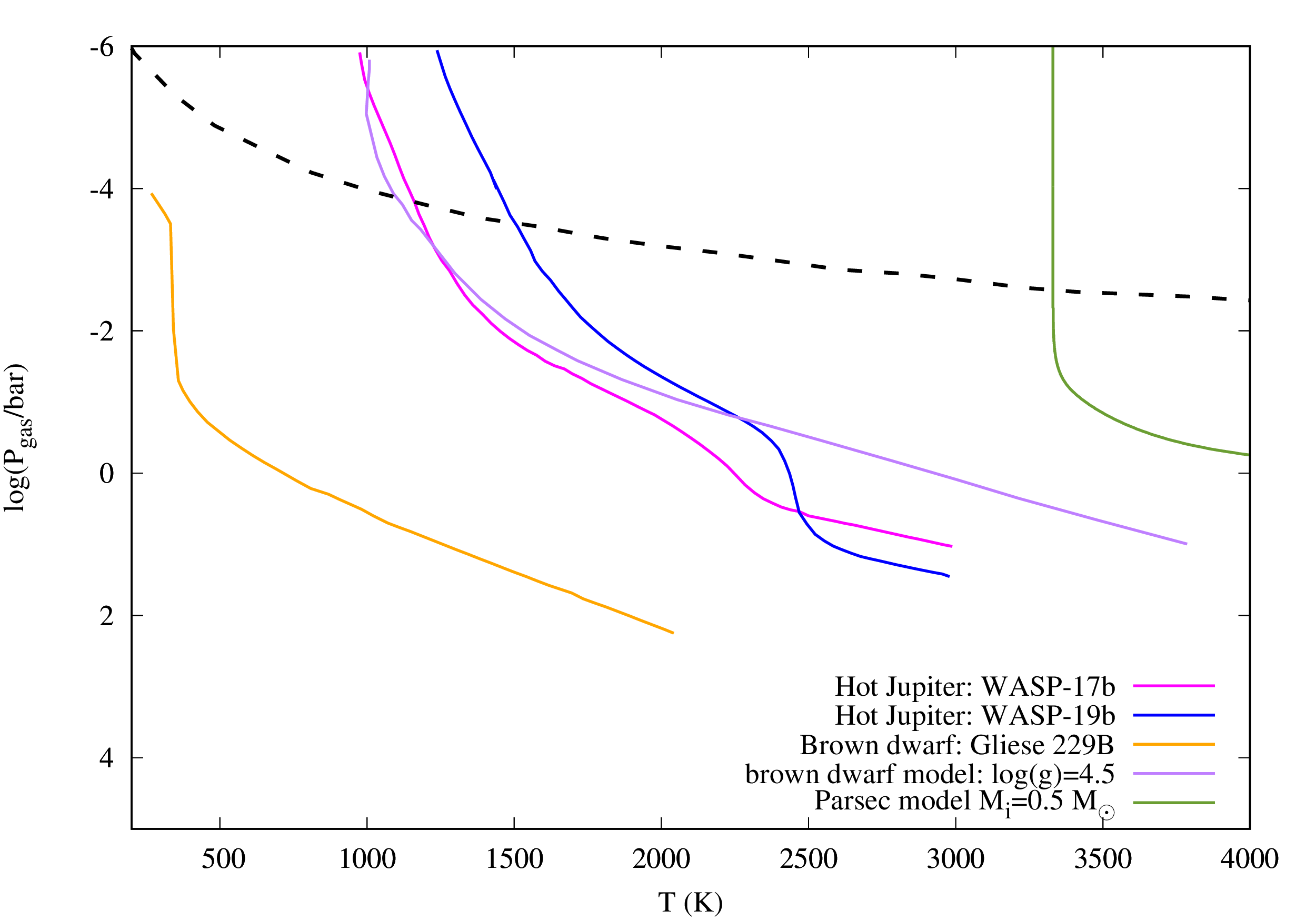} 
    \caption{Atmospheric pressure-temperature profiles for exoplanets, brown dwarfs, and very low mass stars. The two hot Jupiter exoplanets,  WASP-17b with mass of $0.51\,M_{\rm J}$, and WASP-19b with mass of $1.14 \,M_{\rm J}$,  were analyzed by \citet{Sing_etal_16} through HST/Spitzer observations. \citet{Calamari_etal_22} used an atmospheric retrieval analysis to derive the profile for the brown dwarf Gliese 229B with an estimated mass of $50 \,M_{\rm J}$. We also add two theoretical profiles for brown dwarfs with a surface gravity of $\log(g)=4.5$ \citep{Baraffe_etal_02}, and a very low mass star model from \texttt{PARSEC} \citep{Bressan_etal_12}. The dashed black line denotes the limit below which pressure broadening for molecular spectral lines should be included.
    Refer to Section~\ref{ssec_broad} for more details.
} 
\label{fig_PT}
\end{figure}

Rosseland mean opacities are critical components in modeling stars and sub-stellar objects. To our knowledge, the most widely distributed suppliers of Rosseland mean opacities for $T \lesssim 10000$ K are the Wichita State University group \citep{Alexaner_75, AlexanderFerguson_94, Ferguson_etal_05}, and the \texttt{\AE SOPUS} team \citep{Marigo_Aringer_09, aesopus2}. 
Two standard variables are used to build Rosseland mean opacity tables:
\begin{equation}
T\,\,\,\,\,{\rm and}\,\,\,\,\,R=\rho\, T_6^{-3}\,,
\end{equation}
where $T$ is the temperature (in K) and the $R$ parameter (in $\mathrm{g\,cm}^{-3}(10^6\, \mathrm{K})^{-3}$) includes both the temperature ($T_6=T/(10^6\,{\rm K})$) and the gas mass density $\rho$ (in g\,cm$^{-3}$). 
The fact that smooth opacity interpolations are feasible is the primary motivation behind the selection of these two parameters.

So far, low-temperature opacities have been computed encompassing the range  $-8 \leq \log(R) \leq +1$ \citep{Ferguson_etal_05, aesopus2}.
This range, however, is insufficiently broad to cover the structural characteristics of very low-mass stars, brown dwarfs, and planets
(see Section~\ref{sec_eos}).
In fact, these objects have atmospheres with high density and pressure, so $\log(R)$ can easily exceed 1. Very low-mass stars ($0.08 \lesssim M/\Msun \lesssim 0.6$) and sub-stellar objects ($10^{-3} \lesssim M/M_{\rm J} \lesssim 13$ for exoplanets, $13 \lesssim M/M_{\rm J} \lesssim 90$ for brown dwarfs, with $M_{\rm J}\equiv$ 1 Jupiter mass) have atmospheres that typically cover temperature ranges of $100 \lesssim T/{\rm K} \lesssim 3500$ and gas pressure ranges of $10^{-4} \lesssim P_{\rm gas}/{\rm bar} \lesssim 10^3$ \citep{Burrows_etal_01, Spiegel_etal_11, Wilson_etal_16, Mulders_etal_21}. This demonstrates the need of extending the Rosseland mean opacity tables at higher densities and pressures (with $\log(R) > 1$).

The thermal Doppler mechanism is used by both the Wichita State University and \texttt{\AE SOPUS} teams for molecular line broadening. Micro-turbulence velocity is also included in \texttt{\AE SOPUS}. As we will see in Section~\ref{ssec_broad}, this approximation is mostly valid over the standard range $-8 \leq \log(R) \leq +1$, but it becomes inadequate when higher $R$ values, hence larger densities and pressures are considered.
When moving into the high density regime, pressure effects broaden molecular transitions, resulting in either Lorentzian or Voigt profiles \citep{Burrows_etal_01, Sharp_Burrows_07}.
Figure~\ref{fig_PT}  depicts the pressure-temperature atmospheric profiles of two exoplanets, one brown dwarf, and very low mass stars. Pressure broadening of spectral molecular lines affects all of these objects' atmospheres (below the dashed black line).
Few works in the past literature, to our knowledge, computed Rosseland mean gas opacities at high densities, namely \citet{Kurucz_93}, \citet{Freedman_etal_08}, \citet{Malygin_etal_14}, \citet{Freedman_etal_14}, adopting scaled-solar abundances according to \citet{GS_98} or \citet{Lodders_03}. For zero-metallicity gas, we also refer to the work of \citet{Lenzuni_etal_91}.

Under high density conditions, additional important phenomena, namely electron degeneracy and Coulomb interactions between charged particles, must be accounted for in the equation of state \citep{CoxGiuli_68,Hansen_76,Potekhin_etal_09}. Furthermore, all charges present in a radiating gas, electrons and ions, contribute to reduce the energy required to free an electron in the fundamental state.
This process is designated as ionization potential depression \citep[IPD;][]{EckerKroll_63, Stewart_Pyatt_66}. We will show that the IPD effect will have a notable impact on Rosseland mean opacities, especially for $T> 10000$ K.
As a result, appropriate physics must be included to describe the gas at high pressure and its opacity interaction with radiation.

This study aims to increase the accessibility of Rosseland mean gas opacities in high-pressure environments, across a broad temperature range, from 100 K to 32000 K. Expanding the opacities to cover a wide range of pressure and density conditions ($-8 \le \log(R) \le +6$) is essential for accurate modeling of sub-stellar objects and very low mass stars.
We provide our results in a public web interface that allows users to customize the chemical mixture according to their specific research requirements. This flexibility is crucial because different research projects may focus on different chemical compositions and environmental conditions.
In addition we produce pre-computed opacity tables  with the inclusion of condensates, similarly to \citet{marigo23}.

This paper is organized as follows.
Section~\ref{sec_eos} recaps the basic ingredients of the equation of state in \texttt{\AE SOPUS} and \texttt{GGchem} codes, as well the the method for calculating the Rosseland mean opacity. We additionally present some physical structure of very low mass stars to demonstrate the need for the pressure and density ranges of the opacity tables to be expanded to higher values.
Section~\ref{opac} describes how we implement the IPD effect in  \texttt{\AE SOPUS} partition functions and abundance differential equations, as well as how we treat line pressure broadening for atomic and molecular transitions.
In Section~\ref{sec_results}  we analyze and discuss the findings, focusing on the impact of the IPD effect and pressure broadening on Rosseland mean opacities and comparing our results to others found in the literature.
Section~\ref{sec_solidgrains} presents a few examples of Rosseland mean opacities with solid grains included, while Section~\ref{sec_impactmodels} briefly discusses the impact of these opacities in low mass stellar models.

\section{Equation of State and Rosseland mean opacities}
\label{sec_eos}

The \texttt{\AE SOPUS} code solves the equation of state encompassing more than 800 species, including about 300 atoms and ions and 500 molecules, in the gas phase under conditions of thermodynamic and instantaneous chemical equilibrium (see Appendix~\ref{sec_appendix}). In order to expand the opacity computations in the low-temperature regime where solid grains condense, we \citep{marigo23} recently coupled \texttt{\AE SOPUS} \citep{Marigo_Aringer_09, aesopus2} and \texttt{GGchem} codes \citep{ggchem_18}.
In practice, we use \texttt{\AE SOPUS} for temperatures between 30000 K and 3000 K, and then we switch to \texttt{GGchem} for temperatures between 3000 K and 400 K.
We recall that \texttt{GGchem} computes the abundances of approximately 568 gas molecules, 55 liquid species, and nearly 200 types of solid particles. 

There is an extensive description of how to calculate the Rosseland mean opacity in \cite{aesopus2} and  \cite{Marigo_Aringer_09}, so it will not be repeated here.
 To summarize, we compute the total monochromatic opacity cross section per unit mass (in cm$^2$ g$^{-1}$) for any chosen $(\rho, T)$ pair by incorporating all the contributions from true absorption and scattering, for both gas and solid particles.
 
\begin{figure}[h]
    \centering    \includegraphics[width=0.48\textwidth]{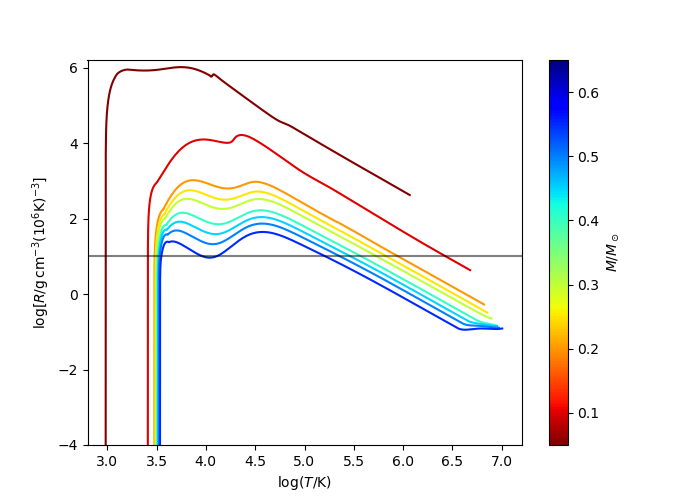} 
    \includegraphics[width=0.48\textwidth]{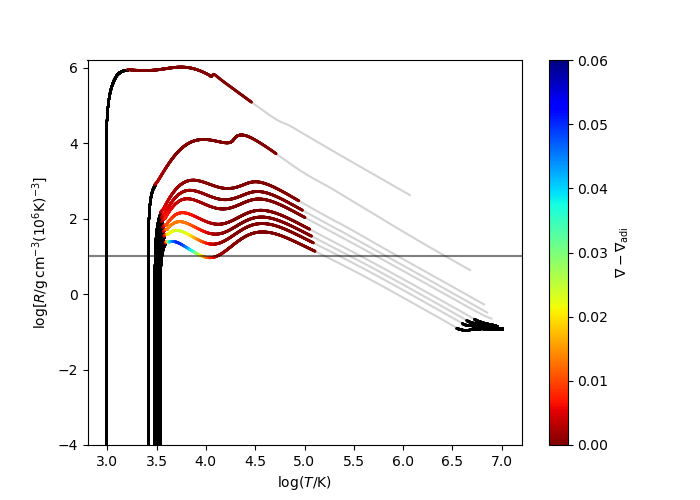} 
    \caption{Structures of very low-mass stars and brown dwarfs  computed with the \texttt{PARSEC} code \citep{Bressan_etal_12}, from the center to the atmosphere, in a stage close to the zero-age main sequence. The initial mass changes in increments of $0.05\,M_{\odot}$,  from $0.05\,\,M_{\odot}$ to $0.7\,\,M_{\odot}$. 
    The initial chemical composition is scaled-solar according to \cite{Caffau_etal_2011}, with a metallicity $Z=0.014$, and helium abundance $Y=0.273$. The horizontal line marks the maximum value,  $\log(R)=1$, currently available in the low-temperature Rosseland mean opacities \citep{Ferguson_etal_05, aesopus2}.
    Left panel: structures in the $\log(T)-\log(R)$ plane, color-coded according to the initial stellar mass. 
    Right panel: structures in the $\log(T)-\log(R)$ plane, with convective regions color-coded according to the degree of super-adiabaticity, $\nabla-\nabla_{\rm ad}$. The gray sections correspond to convective regions treated as adiabatic, while black sections refer to the radiative regions. See text for more details.} 
\label{fig_vlm}
\end{figure}

As previously stated in Section~\ref{sec_intro}, the standard range of Rosseland mean opacities, $-8 \leq \log(R) \leq 1$, is not enough to describe the physical properties of very low-mass stars, brown dwarfs and planets, which reach much higher densities and pressures.
To demonstrate this fact, we display a few stellar structures in Figure~\ref{fig_vlm} that correspond to brown dwarfs and very low-mass stars with initial masses $M_{\rm i}$ between $0.05\,M_{\odot}$ and $0.7\,M_{\odot}$.

Looking at the left panel of Figure~\ref{fig_vlm}, it is evident that a sizable portion of the structures of stars with $M_{\rm i}\le 0.7\,M_{\odot}$ exceeds the upper limit of $\log(R)=1$, reaching values up to $\log(R)\simeq 4$. The brown dwarf model with $M_{\rm i}=0.05 \,M_{\odot}$ extends up to $\log(R)\simeq6$.
As we will discuss in Section~\ref{sec_impactmodels}, opacity extrapolation outside of the validity range of tables may result in incorrect structural properties.

From stellar evolution theory, we know that main-sequence stars with $0.08 \lesssim M_{\rm i}/M_{\odot} \lesssim 0.35$ are completely convective and adiabatic, regardless of the convection model used. However, convection becomes less effective for transporting energy closer to the surface, particularly in layers where hydrogen and helium are partially ionized, and the true temperature gradient, $\nabla=d\log(T)/d\log(P)$, becomes super-adiabatic, that is $\nabla-\nabla_{\rm ad} > 0$. In these regions, micro-physics plays a crucial role, as factors such as the equation of state and opacities become essential for determining the solution of the stellar structure.
The right panel of Figure~\ref{fig_vlm} highlights the regions where the temperature gradient becomes super-adiabatic. The mixing-length theory \citep{MLT_58} is used to estimate $\nabla-\nabla_{\rm ad}$. 
It is clear that in very low-mass stars, low-temperature opacities extend in stellar layers where $\log(R) > 1$ and convection is super-adiabatic, and thus accurate opacity estimation is critical.

\begin{figure}[h]
    \centering    
    \includegraphics[width=\textwidth]{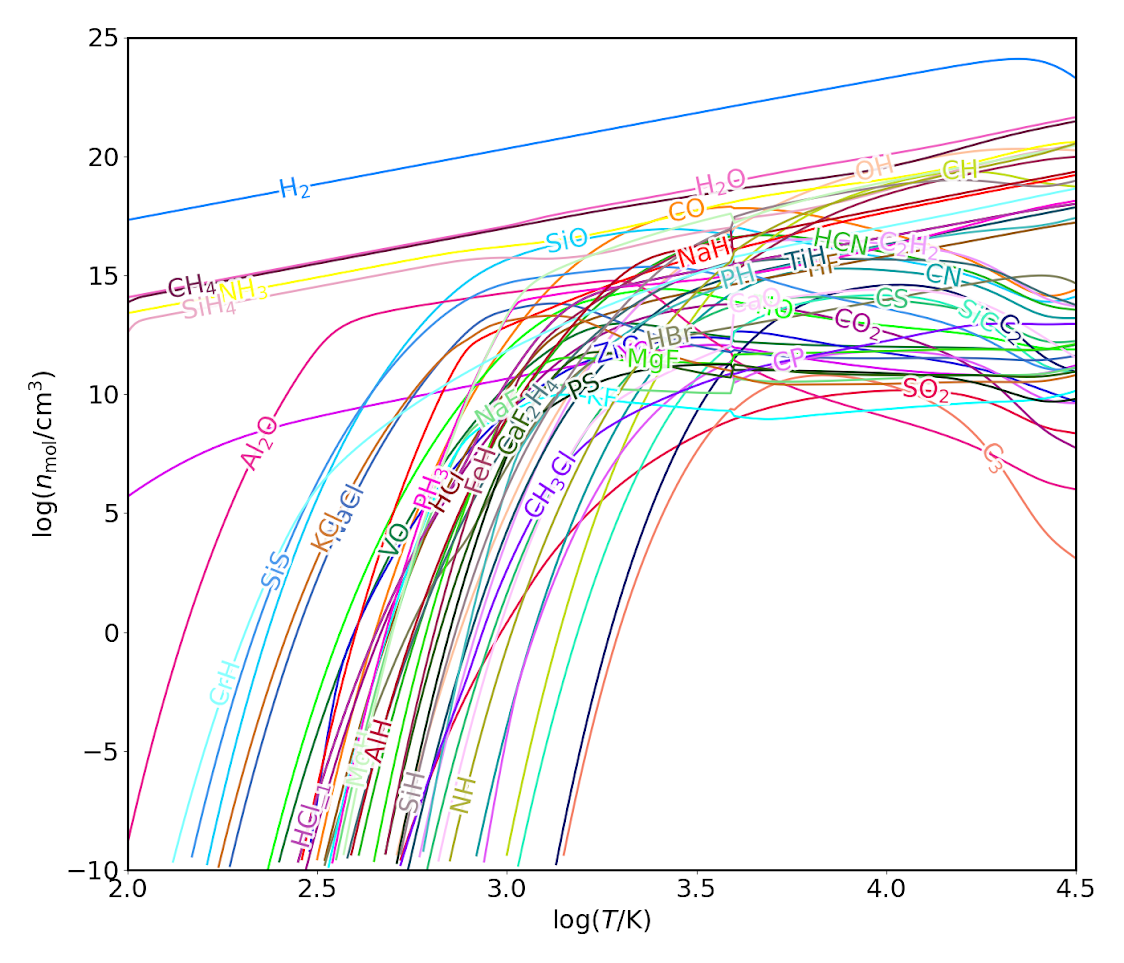}
    \caption{Gaseous molecular concentrations, in number density, for some selected species as a function of temperature, without condensation.
    We assume $\log(R)=6$, with a chemical composition  solar-scaled according to \citet{Magg_etal_22}, with metallicity $Z=0.01$ and hydrogen abundance $X=0.7$.}
\label{fig_molec_cond1}
\end{figure}

\begin{figure}[h]
    \centering    
    \includegraphics[width=\textwidth]{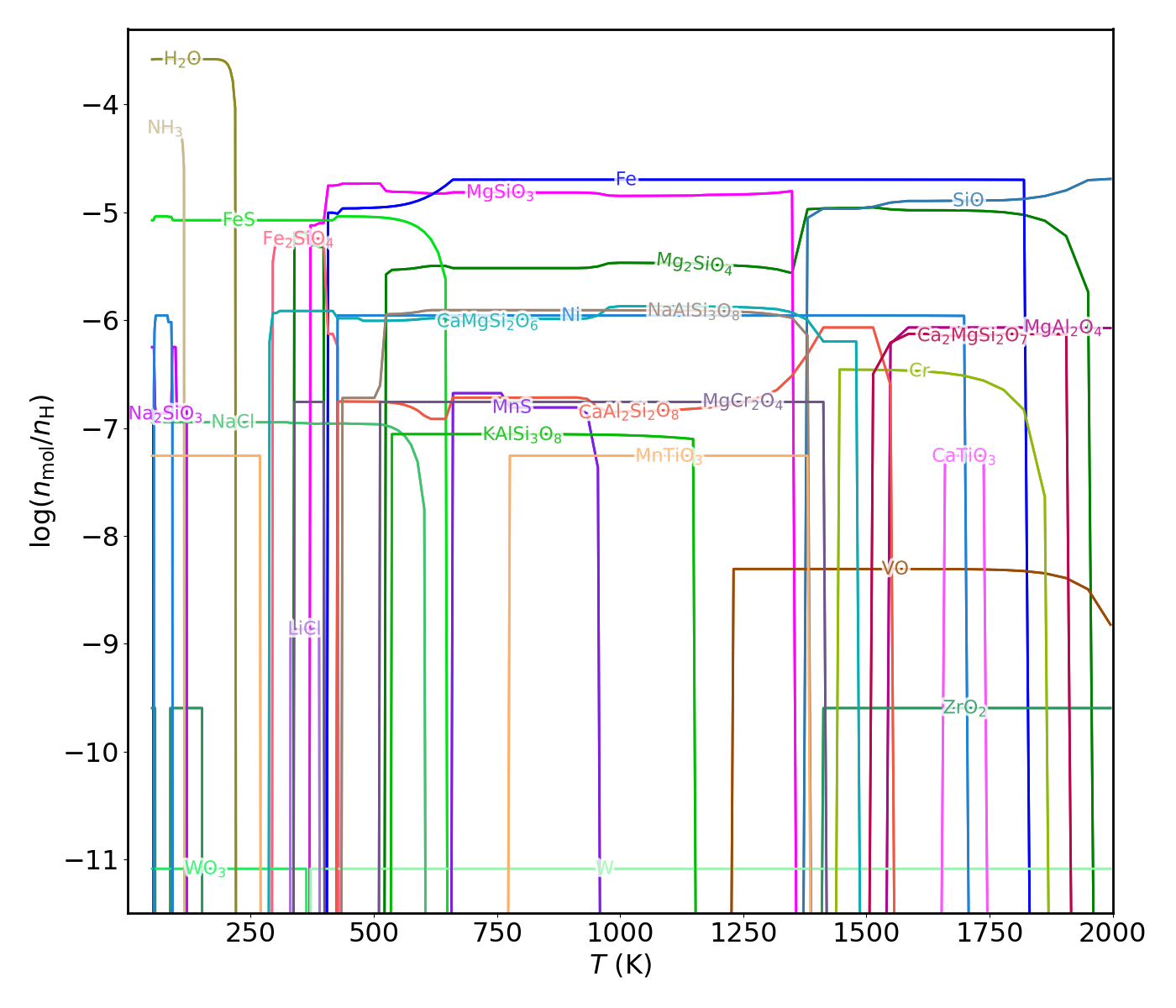}
    \caption{Distribution of several condensed species as a function of temperature at $\log(R)=6$, for solar abundances in phase equilibrium, computed with \texttt{GGchem}. The abundances are evaluated with respect to the number density of hydrogen nuclei. The assumed $\log(R)$ value and chemical composition are the same as in Fig.~\ref{fig_molec_cond2}.}
\label{fig_molec_cond2}
\end{figure}

What are the chemical species that have to be considered in the calculation of Rosseland mean opacities at high values of $R$? To answer this question,
Figure \ref{fig_molec_cond1} depicts the concentrations of several molecules in the gas phase as a function of temperature for $\log(R)=6$. At high density the most abundant molecules are  molecular hydrogen (H$_2$) and water vapor (H$_2$O), carbon monoxide (CO), silicon monoxide (SiO), followed by methane (CH$_4$), ammonia (NH$_3$) and silane (SiH$_4$) towards lower temperatures. 

Figure \ref{fig_molec_cond2} shows a number of condensed species, including metal oxides such as SiO, ZrO, VO, silicates, Cr, Fe. 
It is worth noting that corundum (Al$_2$O$_3$) does not condense in appreciable amounts at high densities (hence it is not shown in Fig.~\ref{fig_molec_cond2}).
The condensation sequence at low temperatures is closed by iron sulfide (FeS), water ice (H$_2$O) and ammonia ice (NH$_3$).

\section{Opacities at high density}
\label{opac}
When entering a high density regime, opacities become increasingly complex and must be handled with great care due to a variety of physical processes, including electron degeneracy, which increases gas pressure; non-ideal effects due to Coulomb interactions among charged particles, which can lower atom and molecule ionization potentials; and line pressure broadening of molecular bands and atomic transitions.
In the following sections, we will detail the necessary improvements we made in \texttt{\AE SOPUS} to deal with these high-density conditions.

\subsection{Ionization potential depression}
\label{ssec_potlow}
\begin{figure}[H]
    \centering    \includegraphics[width=0.48\textwidth]{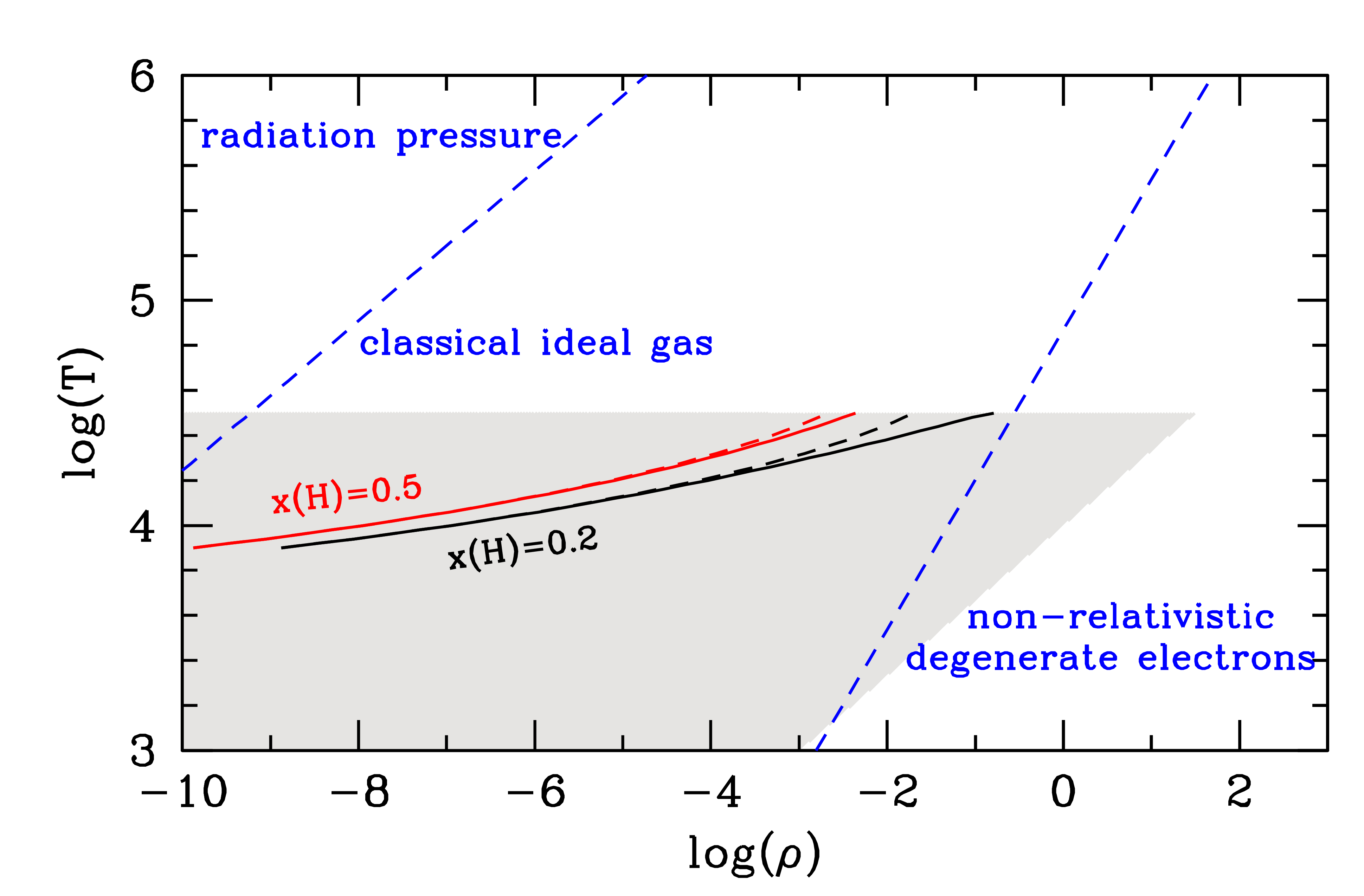}    \includegraphics[width=0.48\textwidth]{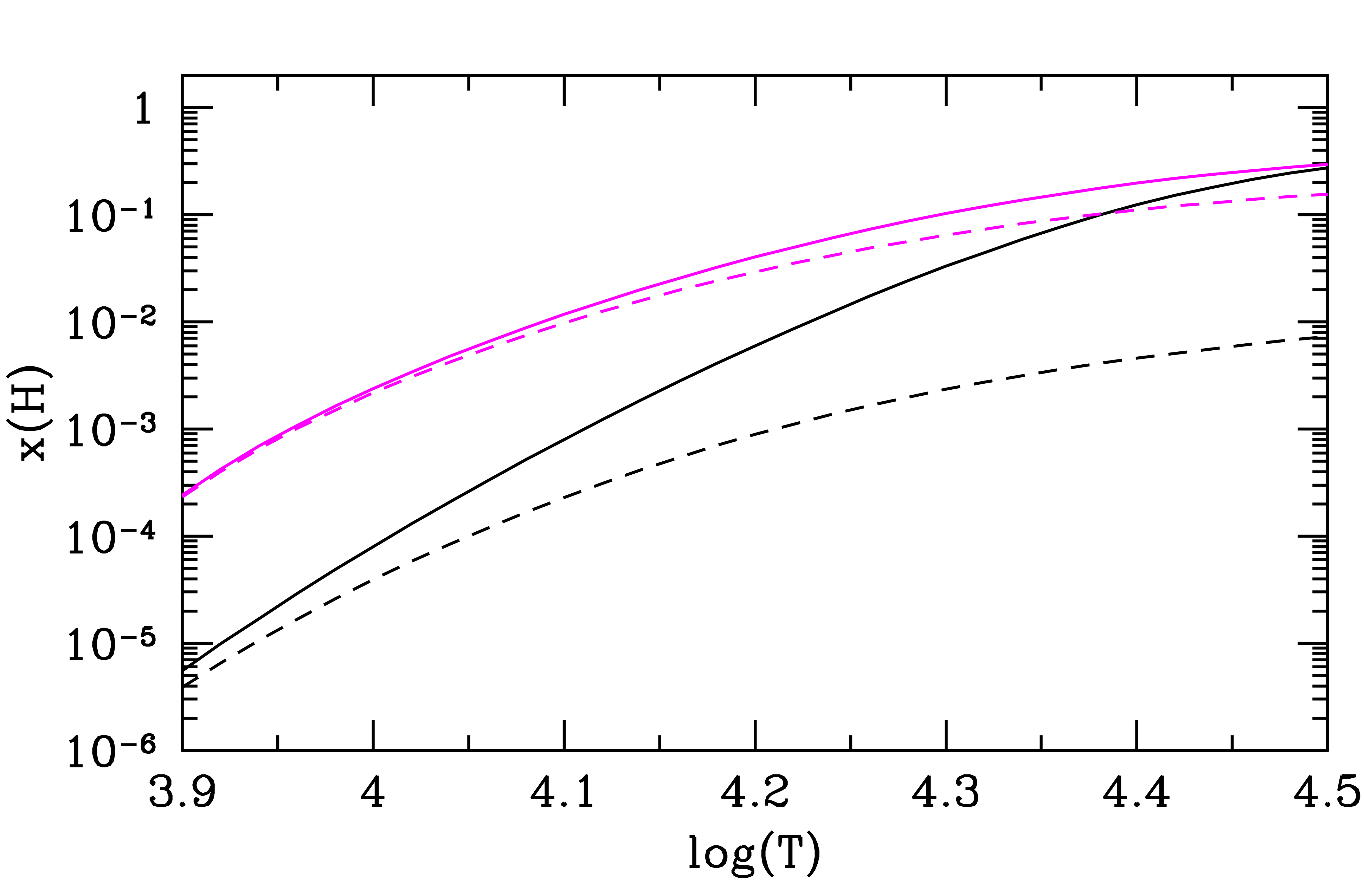}
    \caption{Left panel: The equation of state for a gas of free particles in the $\log(T)-\log(\rho)$ plane. The dashed blue lines are
approximate boundaries between regions where radiation pressure, classical ideal gas pressure, and non-relativistic electron
degeneracy dominate. We assume \cite{Magg_etal_22}'s solar composition with $X = 0.7$ and $Z = 0.01$. The gray area denotes the upper extension of our opacity tables. The red and black lines mark the locus of points where the ionization degree of hydrogen is 0.5 and 0.2, as indicated. 
Right panel: Ionization degree of hydrogen as a function of temperature for $\log(R)=6$ (black line) and $\log(R)=3$ (magenta line).
In both panels the plasma IPD effect is represented by solid lines, while dashed lines are the result of the classical Saha ionization equation. See text for more details. }
    \label{fig_potlow}
\end{figure}
The \texttt{\AE SOPUS} opacity calculations are extended up to $\log(R)=6$ in a high density regime. 
At the highest temperature, $\log(T/\rm K)=4.5$, this corresponds to a mass density of $\rho\simeq 32\, {\rm g\, cm^{-3}}$.
As shown in Figure~\ref{fig_potlow} (left panel), electrons are partially degenerate at this density. We recall that \texttt{\AE SOPUS} incorporates an accurate treatment of electron degeneracy, based on generalized Fermi-Dirac integrals \citep{CoxGiuli_68}.
At increasing densities some non-ideal effects appear, such as the lowering of the ionization potentials of atoms and molecules.
 
Let us briefly discuss this aspect. We know that the ionization potential $U$ of an ion embedded in a plasma is reduced due to the interaction of all charged particles (ions and electrons) with that ion.
The IPD effect is accounted for using the method developed by \citet{EckerKroll_63}, who formulated a generalized Saha equation as a function of the chemical potential of the plasma. 

The IPD is calculated with
\begin{equation}
\Delta U(z) = \frac{z\,e^2}{\lambda_{\rm D}}\, ,
\end{equation}
where $z$ is the charge of the ion after the ionization occurrence, $e$ is the electron's charge, and $\lambda_{\rm D}$ is a generalized Debye length, which is computed through
\begin{equation}
\lambda_{\rm D} =\sqrt{\frac{k T}{4\pi(n_{\rm e}+n_{\rm ion})}}\,,
\end{equation}
where $k$ is the Boltzmann constant, $n_{\rm e}$ and $n_{\rm ion}$ are the number densities of free electron and ions in the plasma.

The implementation of the IPD effect in \texttt{\AE SOPUS} is carried out as follows.
The \texttt{\AE SOPUS} code uses the Newtwon-Raphson technique to solve the equation of state assuming instantaneous chemical equilibrium, employing a set of dissociation-recombination and/or ionization equilibrium constants for each of the 800 particles.
Concerning the Saha equation for ionization, at each iteration we correct the ionization potential using the IPD effect inside the equilibrium constant, which depends on $U(z)-\Delta U(z)$,  $T$, $n_{\rm e}$ and $n_{\rm ion}$, until convergence is reached.

This is illustrated in the right panel of Figure~\ref{fig_potlow}. It compares the hydrogen ionization degree $x({\rm H})$\footnote{The ionization degree of hydrogen is defined as the ratio $\mathrm{x(H)= n(H^+)/[n(H)+n(H^+)]}$, where $\mathrm{n(H)}$ and $\mathrm{n(H)^+}$ are
the number densities of neutral and ionized hydrogen.
} with and without the IPD effect as a function of temperature for two $R$ parameter values. Particularly for $\log(R)=6$, with the classical Saha equation, $x({\rm H})$ remains extremely low, whereas with the IPD plasma effect, $x({\rm H})$ increases significantly at the highest temperatures.

The locus of points in the $\log(T)-\log(\rho)$ diagram where the ionization degree for hydrogen $x({\rm H})$ is equal to 0.5 and 0.2 is plotted in Figure~\ref{fig_potlow} (left panel).
We observe that, when the IPD effect is considered, the same ionization degree is obtained at higher densities than in the ideal case. This is especially noticeable at the highest temperatures for $x({\rm H})=0.2$, while at lower densities the differences become less and less significant.

The reduction in the ionization potential of atoms and molecules has a sizable impact on the Rosseland mean opacities at high density, as discussed in Section~\ref{ssec_khighro}.

\subsection{Line pressure broadening for atomic and molecular transitions}
\label{ssec_broad}
Various processes in planet, substellar, and stellar atmospheres naturally broaden spectral lines.  Doppler and pressure broadening are the most common types of line broadening. 
Doppler broadening is caused by the thermal velocities of each atom and molecule and is normally described by a Gaussian line profile \citep{EXOCROSS_2018}. The width of the Doppler line core is directly related to the temperature. As the temperature increases, the thermal motion of particles becomes more significant, causing broader line profiles.

For instance, \cite{Ferguson_etal_05}  uses a pure thermal Doppler profile in their opacity computations. In \texttt{\AE SOPUS} we do similarly but also account for micro-turbulence velocity by producing a normalized broadening profile, $\phi(\nu)$, according
to the equation:
\begin{equation}
\phi(\nu)=\frac{1}{\Delta_{\nu}\sqrt{\pi}}\,e^{-\left(\frac{\nu-\nu_0}{\Delta_{\nu}}\right)^2}\,,
\end{equation}
where $\nu_0$ is the line center position in frequency, and $\Delta_{\nu}$ is the line width, obtained with
\begin{equation}
\Delta_{\nu}=\frac{\nu_{0}}{c}\sqrt{\frac{2 k_{\rm B}T}{m}+\xi^2}.
\label{eq_dopplermicro}
\end{equation}
In this Equation, $c$ denotes the speed of light, $k_{\rm B}$ the Boltzmann constant, $m$ the molecule's mass, and $\xi$  the micro-turbolent velocity, which is set to $2.5$~km/s \citep[see][for more details]{Marigo_Aringer_09, aesopus2}.
\startlongtable
\begin{deluxetable}{clccccc}
\tabletypesize{\footnotesize}
\tablecolumns{7}
\tablecaption{Spectral Line Data for Molecular Absorption with Pressure Broadening taken from the \texttt{ExoMolOP} Database \label{tab_exomolop}}
\tablehead{
\colhead{Group} & \colhead{Species}	&	\colhead{Line list}	&	\colhead{Reference}	&	\colhead{$\lambda_{\rm l}$} 	&	\colhead{$\lambda_{\rm u}$} &	\colhead{$T_{\rm max}$}\\
 & &		&	 &	 \colhead{($\micron$)}	&	 \colhead{($\micron$)}	&	 \colhead{(K)} 
}
\startdata
\multicolumn{1}{l}{Metal Oxides} \\
		\rule{0pt}{3ex} & AlO	&	ExoMol ATP	&	\cite{AlO_2021MNRAS.508.3181B}	&	0.29	&	100	&	8000		\\
			& CaO	&	ExoMol VBATHY	&	\cite{CaO_10.1093/mnras/stv2858}	&	0.5	&	100	&		5000	\\
			& MgO	&	ExoMol LiTY	&	\cite{MgO_10.1093/mnras/stz912}	&	0.3	&	100	&	5000 \\
			& SiO	&	ExoMol EBJT	&	\cite{SiO_2022MNRAS.510..903Y}    	&	1.65	&	100	&	9000	\\
			& TiO	&	ExoMol TOTO	&	\cite{TiO_10.1093/mnras/stz1818}	&	0.33	&	100	&	5000\\
			& VO	&	ExoMol VOMYT	&	\cite{VO_10.1093/mnras/stw1969}	&	0.29	&	100	&		5000 \\
\multicolumn{1}{l}{Other Oxides}\\
			\rule{0pt}{3ex} & CO	& Li 2015 		&	\cite{CO_Li_2015}	&	0.43	&	100	&	5000 \\ 
			& NO	&	 HITEMP-2019 	& \cite{NO_10.1093/mnras/stab1154} 		&	0.37	&	100	& 4000 \\ 
			& O$_2$	&	HITRAN	&	\cite{O2_2021AA...646A..21C}	&	1.43		&	100	&	296	\\
			& PO	&	ExoMol POPS	&	\cite{PO_PS_10.1093/mnras/stx2229}	&	0.83	&	100	&	5000\\
\multicolumn{1}{l}{Triatomics}\\   
			\rule{0pt}{3ex} & CO$_2$	&	ExoMol UCL-4000	&	 \cite{CO2_10.1093/mnras/staa1874}	& 0.5	& 100 	& 4000	\\
			& H$_2$O	&	POKAZATEL	&	\cite{H2O_10.1093/mnras/sty1877}	&	0.24	&	100	&	4000		\\
			& H$_2$S	&	ExoMol AYT2	&	\cite{H2_2019AA...630A..58R}		&	0.91	&	100	&	2000\\
		&	HCN	&	ExoMol Harris	&	\cite{HCN_10.1111/j.1365-2966.2005.09960.x}	&	0.56	&	100		&	4000\\
		&	O$_3$ &	HITRAN 	&	\cite{HITRAN2020_GORDON2022107949}	&	1.43	&	100 &	-		\\
		&	SiH$_2$ & ExoMol CATS & \cite{SiH_10.1093/mnras/stx2738}  & 1.00 &100   & 2000 \\
		&	SiO$_2$ & ExoMol OYT3 & \cite{SiO2_10.1093/mnras/staa1287}  &1.67  &100 & 3000\\
		&	SO$_2$	&	ExoMol ExoAmes 	&	\cite{O2_2021AA...646A..21C}	&	1.25	&	100		&	2000\\ 
\multicolumn{1}{l}{Metal hydrides}\\
			\rule{0pt}{3ex} & AlH	&	ExoMol AlHambra	&	\cite{AlH_2018MNRAS.479.1401Y}	&	0.37	&	100	&	5000	\\
		&	BeH	&	ExoMol 	&	\cite{BeH_Darby_Lewis_2018}	&	0.24	&	100	&	2000\\
		&	CaH 	&  MoLLIST & \cite{CaH_MgH_10.1093/mnras/stac371}	&	0.45	&	100	& 5000	\\
		&	CrH	&	MoLLIST	&	\cite{Bernath_etal_20}	&	0.69	& 100		&	5000\\
		&	FeH	&	MoLLIST	&	\cite{FeH_Dulick_2003}	&	0.67	&	100	&	5000	\\
		&	LiH	&	CLT &	\cite{LiH_2011MNRAS.415..487C} 	&	0.5	&	100	&	2000	\\
		&	MgH 	&	MoLLIST	&	 \cite{MgH_CaH_2022MNRAS.511.5448O}	&	0.34	&	100	&		2000	\\
	&		NaH	& ExoMol Rivlin		&	\cite{NaH_2015MNRAS.451..634R}	&	0.27	&	100	&	7000	\\
		&	ScH	&	LYT	&	\cite{ScH_2021AA...646A..21C} 	&	0.63	&	100	&	2000	\\
		&	TiH	&	MoLLIST	&	\cite{TiH_Burrows_2005}	& 0.42		&	100	&	5000\\
\multicolumn{1}{l}{Other hydrides}\\
\rule{0pt}{3ex} & CH	&	MoLLIST	& \cite{CH_Masseron_etal_2014}	&	0.26	&	100	&	5000	\\
		&	HBr & HITRAN &	\cite{Li_etal_13}		& 0.62		& 100	&	5000	\\
		&	HCl	&	HITRAN 	&	\cite{Li_etal_13}	&	0.49	&	100	& 5000	\\
		&	HF	&	HITRAN 	&	\cite{Li_etal_13}	& 0.31	&	100	&	5000		\\
		&	HI &  HITRAN &	\cite{Li_etal_13}	&	0.71	&	100	&	5000	\\
		&	NH & MoLLIST & \cite{NH_FERNANDO201829} 	&	0.59	&	100	 		&	5000	\\
		&	OH	& MoLLIST		&	\cite{OHp_Hodges_2017} 	&	0.23	&	100	&	5000	\\
		&	PH	&	ExoMol LaTY	&	\cite{PH_10.1093/mnras/stz1856}		&	0.41	&	100	&	4000\\
		&	SiH	&	ExoMol  SiGHTLY	&	\cite{SiH_10.1093/mnras/stx2738}	&	0.32	&	100	&	5000	\\
		&	SH	& ExoMol GYT &	\cite{SH_10.1093/mnras/stz2517}	& 0.26	&	100	&	5000	\\
\multicolumn{1}{l}{Other diatomics}\\
		& AlCl	&	MoLLIST	&	\cite{Bernath_etal_20}		&	4.26	&	100	& 5000	\\
		&	AlF	&	MoLLIST	&	\cite{Bernath_etal_20}	&	2.58	&	100	&	5000	\\ 
		&	C$_2$	&	 ExoMol 8states	&	\cite{C2_10.1093/mnras/sty2050}	&	0.21	&	100	&	5000	\\
		&	CaF	&	MoLLIST	&	\cite{CaF_HOU201844}	&	1.79	&	100	&	5000	\\  
		&	CN	&	MoLLIST	&	\cite{CN_10.1093/mnras/stab1551}	&	0.23	&	100	&	5000	\\
		&	CP	&	MoLLIST	&	\cite{CP_QIN2021107352}		&	0.67	&	100		&	5000	\\
		&	CS	&	ExoMol JnK	&	\cite{CS_10.1093/mnras/stv1543}	&	0.91	&	100		&	3000	\\ 
		&	H$_2$	& RACPPK &	\cite{H2_2019AA...630A..58R} 	&	0.28	&	100	&	5000	\\
		&	KCl	&	ExoMol Barton	&	\cite{NaCl_KCl_10.1093/mnras/stu944}	&	3.45	& 100		&	3000	\\
		&	KF	&	MoLLIST	&	\cite{NaF_KF_FROHMAN2016104}	&	2.49	&	100	&	5000	\\
		&	LiCl	&	MoLLIST	&	\cite{LiCl_Bittner_2018}	&	2.07	&	100	&	5000	\\
		&	LiF	&	MoLLIST	&	\cite{LiF_Bittner_2018}	&	5.52	&	100	&	5000	\\ 
		&	MgF	&	MoLLIST	&	\cite{MgF_HOU2017511}	&	1.83	&	100		&	5000	\\ 
		&	NaCl	&	ExoMol Barton	&	\cite{NaCl_KCl_10.1093/mnras/stu944}	&	4.00	& 100	&	3000 	\\
		&	NaF	&	 MoLLIST	&	\cite{NaF_FROHMAN2016104}	& 2.01	&	100		&	5000	\\ 
		&	NS	&	ExoMol SNaSH	&	\cite{NS_10.1093/mnras/sty939}	&	0.26	&	100	&	5000 \\
		&	PN	& ExoMol YYLT	&	\cite{PN_10.1093/mnras/stu1854}	&	1.54	&	100		&	5000	\\ 
		&	PS	&	ExoMol POPS	&	\cite{PO_PS_10.1093/mnras/stx2229}	& 0.27		&	100		&	5000	\\ 
		&	SiS	&	ExoMol UCTY	&	\cite{SiS_10.1093/mnras/sty998}	&	2.70	&	100		& 5000	\\ 
\multicolumn{1}{l}{Larger molecules}\\
		&	C$_2$H$_2$	&	ExoMol aCeTY	&	\cite{C2H2_10.1093/mnras/staa229}	&		1.00	&	100	&	2200 \\
		&	C$_2$H$_4$	& ExoMol MaYTY		&	\cite{C2H4_10.1093/mnras/sty1239}	&1.41	&	100	& 700\\ 
		&	CH$_3$ &	ExoMol AYYJ	&	\cite{CH3_2019JPCA..123.4755A}	&	1.00 &	100		&		1500		\\
	&		CH$_3$Cl	&	ExoMol OYT	&	\cite{CH3CL_10.1093/mnras/sty1542}	&	1.56	&	100	&	1200	\\
		&	CH$_4$	&	ExoMol 34to10	&	\cite{CH4_2017AA...605A..95Y}	& 0.56		& 100 & 2000	\\ 
		&	NH$_3$	&	ExoMol CoYuTe	&	\cite{NH3_ALDERZI2015117}	&	0.5	&	100	&	1500		\\
		&	PH$_3$ &	ExoMol SAlTY	&	\cite{PH3_10.1093/mnras/stu2246}	&	1.00	&	100	&	1500	\\
		&	SiH$_4$	&	ExoMol OY2T	&	\cite{SiH4_10.1093/mnras/stx1952}	&	2.00	&	100		&	1200	\\
\multicolumn{1}{l}{Molecular ions}\\
			& H$_3$$^{+}$	&	ExoMol MiZATeP	&	\cite{H3p_10.1093/mnras/stx502}	&	0.4	&	100		&	5000		\\
		&	H$_3$O$^{+}$	&	ExoMol eXeL &	\cite{H3Op_10.1093/mnras/staa2034}	&	1.00	&	100	&	1500		\\
		&	HeH$^{+}$ &	ADJSAAM	&	\cite{HeHp_Amaral_2019}	&0.67	&	100	&	4000 \\ 
		&	LiH$^{+}$	&	CLT	&	\cite{LiHp_10.1111/j.1365-2966.2011.18723.x}	&	10.87	&	100		&	2000\\ 
		&	OH$^{+}$	&	MoLLIST	&	\cite{OHp_Hodges_2017}	&		0.33	&	100	&	5000	\\
\multicolumn{1}{l}{Alkali neutral atoms}\\
		&	K	&	NIST	&	\cite{NISTWebsite}		&	0.29	&	100		&	5000\\
		&	Na	&	NIST	&	\cite{NISTWebsite}		&	-	&	-	&	5000\\ 
\multicolumn{1}{l}{Other atoms}\\
& \multirow{5}{0cm}{1}{C N O} & & & & &   \\
& {Ne Na Mg} & & & & & \\
& {Al Si S} &  Opacity Project & \cite{Seaton_etal_94}	&	-	&	-	&	$10^8$\\
& {Ar Ca Cr} & & & & & \\
& {Mn Fe Ni} & & & & & \\
\enddata
\tablecomments{For each species $\lambda_{\rm l}$, $\lambda_{\rm u}$ denote the minimum and maximum wavelength of the corresponding line list; $T_{\rm max}$ is the highest temperature available.}
\end{deluxetable}

Pressure broadening, which varies depending on the perturbing species (such as H, He, H$_2$) in addition to pressure, results in a Lorentzian or Voigt profile. 
While a Lorentzian profile is typically pressure-dependent,
a Voigt profile is a convolution of the Doppler and Lorentzian profiles. It accounts for both thermal motion and collisional broadening, making it suitable for modeling line shapes in a broader range of conditions.

In addition to temperature and pressure, the line width in Lorentzian and Voigt profiles is also influenced by various other quantities, the so-called broadening parameters, that can be challenging to determine accurately.
Broadening parameters, if available, are present in \texttt{ExoMol} database  \citep{Barton_etal_17, Yurchenko_etal_17}.
Based on \texttt{ExoMol} broadening data, \cite{Chubb_etal_21} recently created a publicly accessible database (\texttt{ExoMolOP}\footnote{Data is available at \url{https://www.exomol.com}}) of opacities for over 80 molecules of astrophysical interest computed at various pressures ($10^{-5}$ to $10^{2}$ bar) and temperatures (the range depends on the line list). Atomic data for the alkali neutral metals, Na and K, is additionally provided, based on \texttt{NIST} database \citep{NIST_ASD}  and the most recent profiles for the resonance lines \citep{Allard_etal_16, Allard_etal_19}.
The data can be recovered in a variety of formats that are compatible with different exoplanet atmosphere retrieval codes.

For this work we use the cross section data for the retrieval code \texttt{TauREx} \citep{Waldmann_etal_15a, Waldmann_etal_15b, Al-Rafaie_etal_21}, in HDF5 format, with a spectral resolution of $R=\lambda/\Delta \lambda=15000$, wavenumber coverage of $200-33333$ cm$^{-1}$.
The \texttt{TauREx} table format is compatible with another retrieval code \texttt{petitRADTRANS} \citep{Molliere_etal_19},  so we could benefit from  
its publicly available software.
Utilizing the \texttt{Exo\textunderscore k}\footnote{\url{https://perso.astrophy.u-bordeaux.fr/~jleconte/exo_k-doc/index.html}} code \citep{Leconte_etal_21}, 
we can create cross section tables appropriate for \texttt{\AE SOPUS} that are interpolated in pressure for each temperature value.
The grids of temperature and pressure typically have $18$ and $15$ nodes, respectively, distributed throughout the corresponding ranges (the number of temperature nodes as well as the wavenumber grid may vary depending on the line list).
Table~\ref{tab_exomolop} contains a complete list of molecules, alkali neutral metals, and other atoms for which we have pressure broadening line profiles.
For atoms with temperatures above 4000 K, we use \texttt{Opacity Project} cross sections \citep{Seaton_etal_94}, which are expressed as a function of temperature and electron density.
 Line broadening includes effects produced by thermal Doppler, radiation damping and pressure.
 
The molecular data consist of 65 species, which is slightly less than the 80 species included in \citet[][see their Table 2]{aesopus2}. The molecules that currently lack pressure broadening are: O$_3$, ClO, HI, CS$_2$, OCS, NaO, N$_2$, KOH, H$_2$, HCl, ZrO,  C$_3$, CH$_3$Cl, SO. We will incorporate new data into \texttt{\AE SOPUS} as soon as it becomes available.
To avoid opacity gaps, we include monochromatic cross sections for these molecules using thermal Doppler plus micro-turbulent velocity profiles for any ($T$,$R$) combinations.
We interpolate the \texttt{TauREx} tables  as a function of wavenumber, temperature, and pressure to compute monochromatic cross sections for molecules, and alkali atoms Na and K. The same procedure is applied for all molecules at $T$ values above the $T_\mathrm{max}$ limits listed in Table~\ref{tab_exomolop}. 

\begin{figure}[h!]
    \centering    \includegraphics[width=0.50\textwidth]{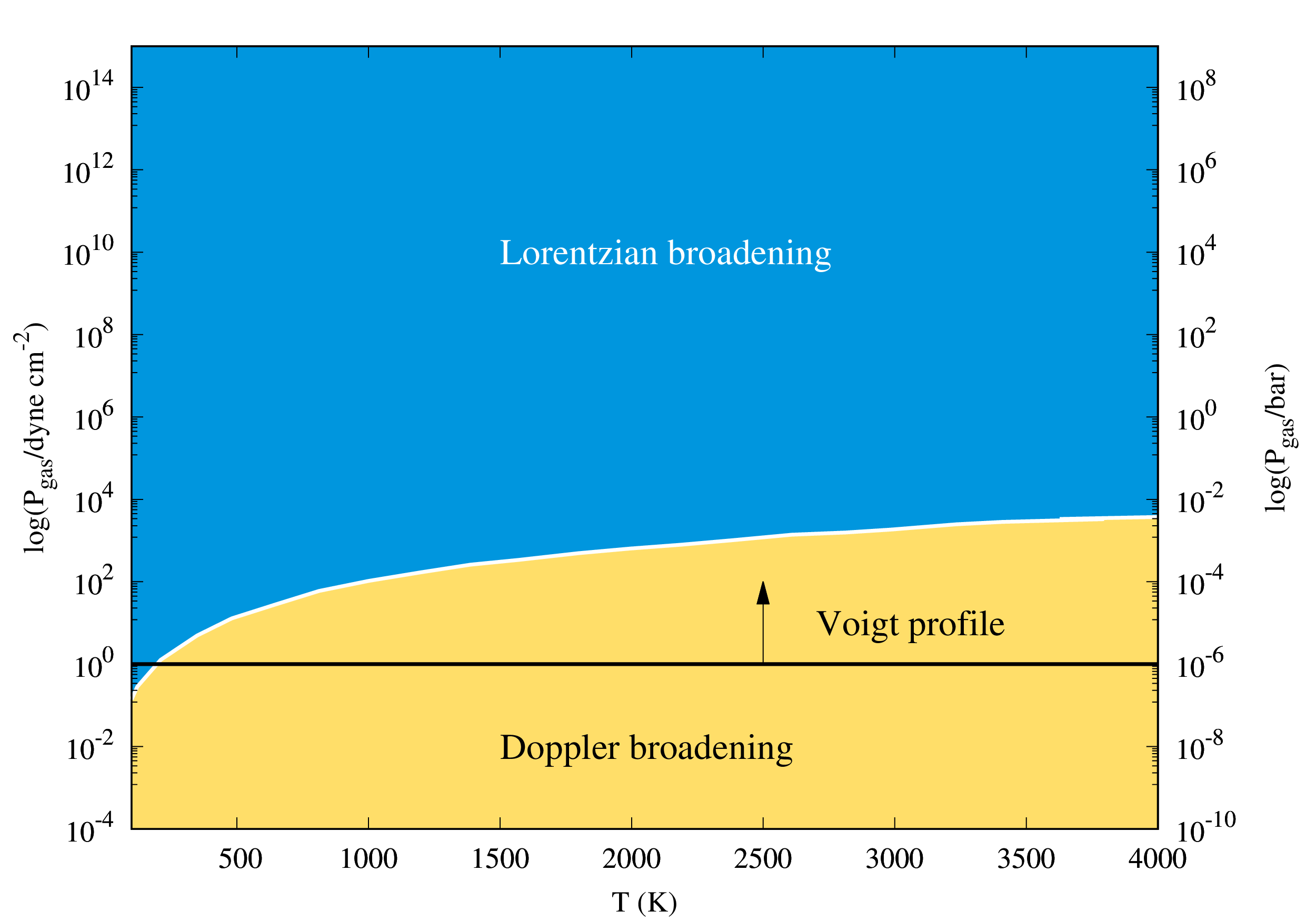}\caption{Comparison  of the widths of Doppler line cores versus Lorentzian profiles in the pressure-temperature diagram according to the analysis of \citet{Hedges_Madhusudhan_16}.
    The boundary between the two regimes is denoted by the white line. Above the  horizontal black line  molecular transitions are treated with a Voigt profile extracted from the \texttt{ExoMolOP} database \citep{Chubb_etal_21}.}
    \label{fig_boundary}
\end{figure}

It is important to know where in the pressure-temperature ($P-T$) space each of the two broadening mechanisms contributes most significantly.
\citet{Hedges_Madhusudhan_16} compared Doppler and Lorentzian broadening profiles over the $P-T$ diagram in terms of half-width at half-maximum to gain a picture of where each profile impacts substantially. 
Based on their findings, we depict the two broadening regimes in Figure~\ref{fig_boundary}. According to \citet{Chubb_etal_21} analysis, we also add a lower pressure limit ($10^6$ bar, black line) above which molecular lines are treated with a Voigt profile. The data can be found in the \texttt{ExoMolOP} database.
As expected, thermal Doppler broadening contributes significantly to the final profile core at low pressures, whereas pressure (Lorentzian) broadening is more effective at high pressures.
Both broadening mechanisms are likely to contribute considerably to the core of the line profiles closer to the border between these two regimes. 
In this work, we use the Voigt profile for molecular lines, which is a convolution of Lorentzian and Doppler broadening mechanisms.
This convolution takes into account both pressure broadening (Lorentzian) and temperature-induced broadening (Doppler), making it a versatile tool for accurately modeling spectral lines in a variety of physical conditions.
Below the black horizontal line in Figure~\ref{fig_boundary} we assume that pressure effects become insignificant, and we use the thermal Doppler plus micro-turbulence velocity broadening.

\begin{figure}[h!]
    \centering    \includegraphics[width=0.48\textwidth]{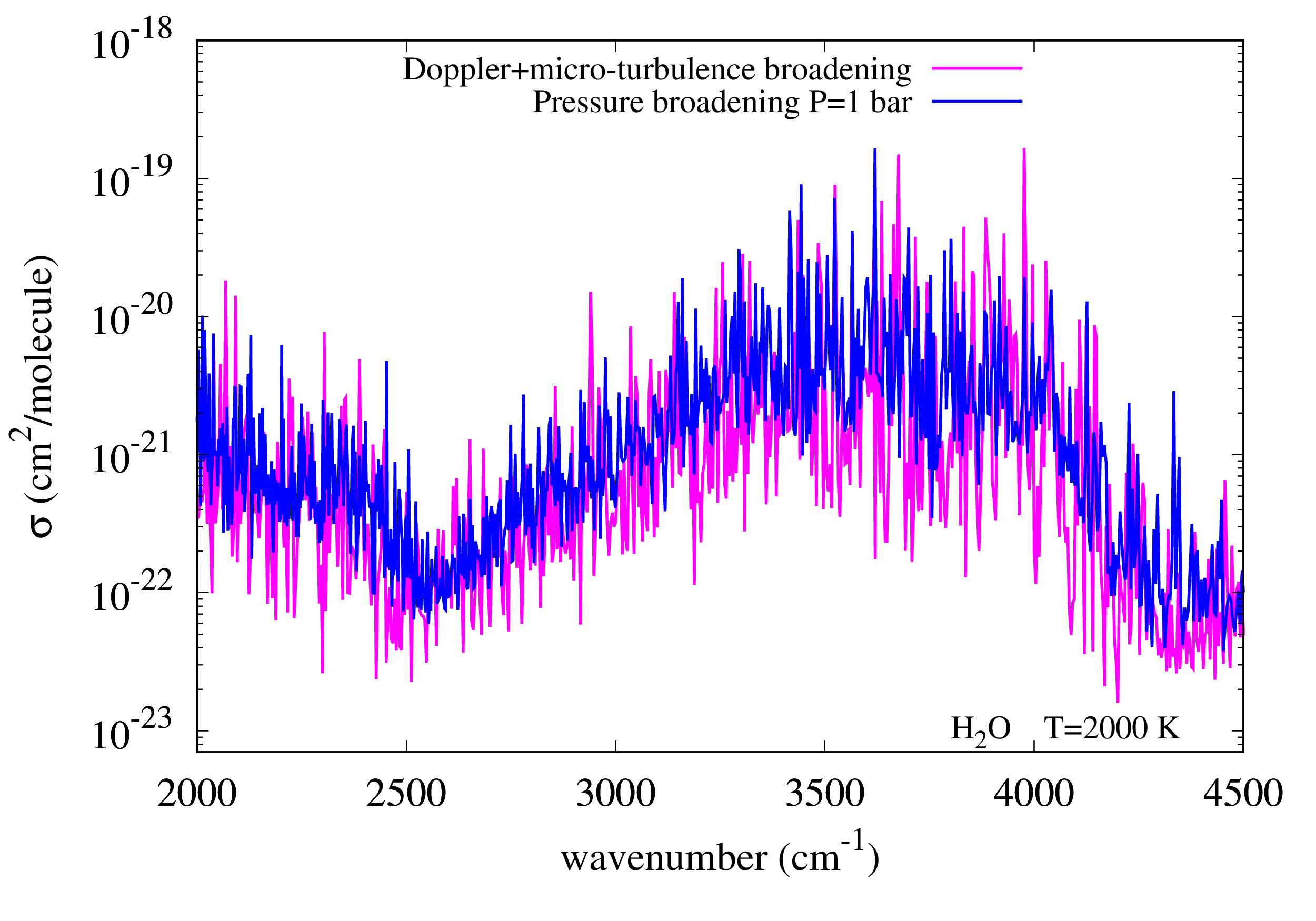}    \includegraphics[width=0.48\textwidth]{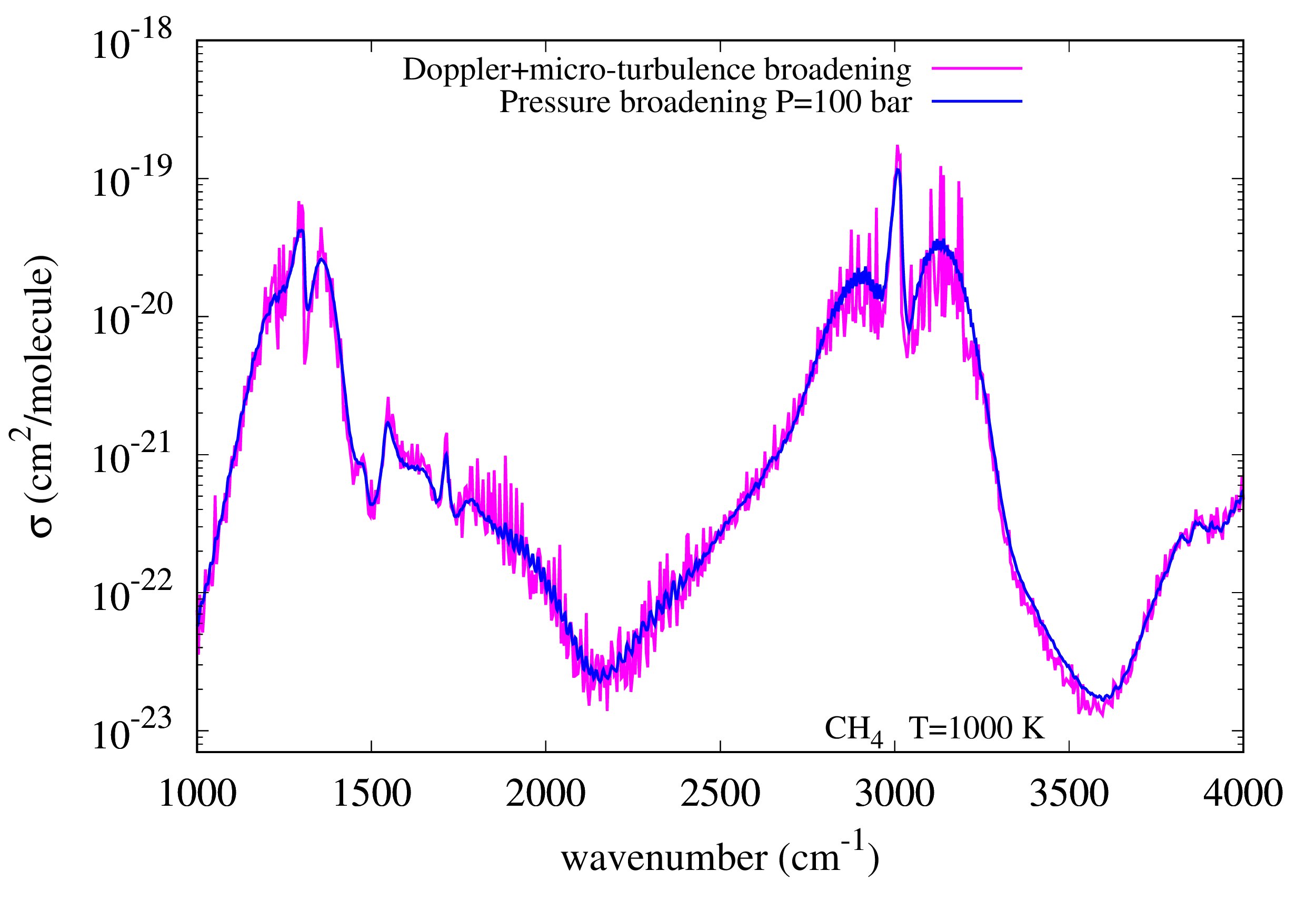}
\caption{Comparison of two broadening profiles used for monochromatic absorption cross sections: the Lorentzian profile produced by pressure and perturbing species (blue line) and the Gaussian thermal Doppler plus micro-turbulent velocity profile (magenta line; see Equation~\ref{eq_dopplermicro}). Left panel: water vapor; right panel: methane. See Table~\ref{fig_table_broad} for details. Values for pressure and temperature are labeled.}
    \label{fig_broadening}
\end{figure}

High pressure can have a sizable impact on the monochromatic absorption cross sections $\sigma$.
 To illustrate the effect Figure~\ref{fig_broadening} compares  $\sigma$ with applied Doppler plus micro-turbulence broadening to pressure broadening, for two molecules, water vapor (H$_2$O) and methane (CH$_4$).
It is evident that pressure broadening reduces the excursion of $\sigma$ to higher and lower values. In the case of water vapor, this is particularly clear at 2000 K. When compared to the thermal Doppler Gaussian profiles, the Lorentzian line profiles produce a sigma that is most concentrated at higher values at a pressure of 1 bar. At 100 bar of pressure and 1000 K of temperature, methane experiences a similar effect, with a cross section that does not exhibit large fluctuations when compared to the Doppler line profiles.
The tendency of $\sigma$ towards higher values at increasing pressure will have a noticeable impact on Rosseland mean gas opacities, which will tend to increase.

\section{Results}
\label{sec_results}
\subsection{Ionization potential depression effects}
 \label{ssec_khighro}
 \begin{figure}[h!]
    \centering    \includegraphics[width=0.48\textwidth]{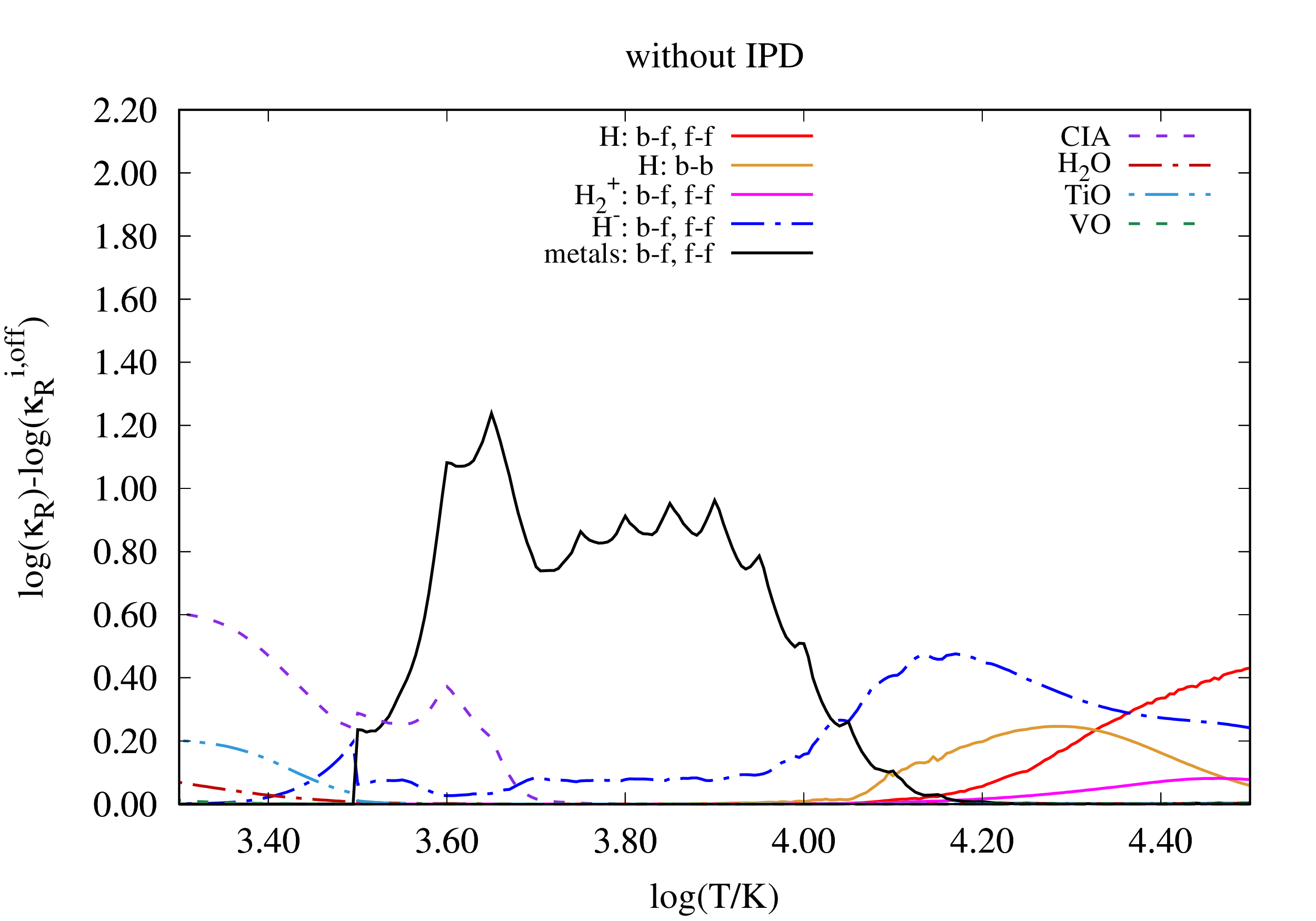}    \includegraphics[width=0.48\textwidth]{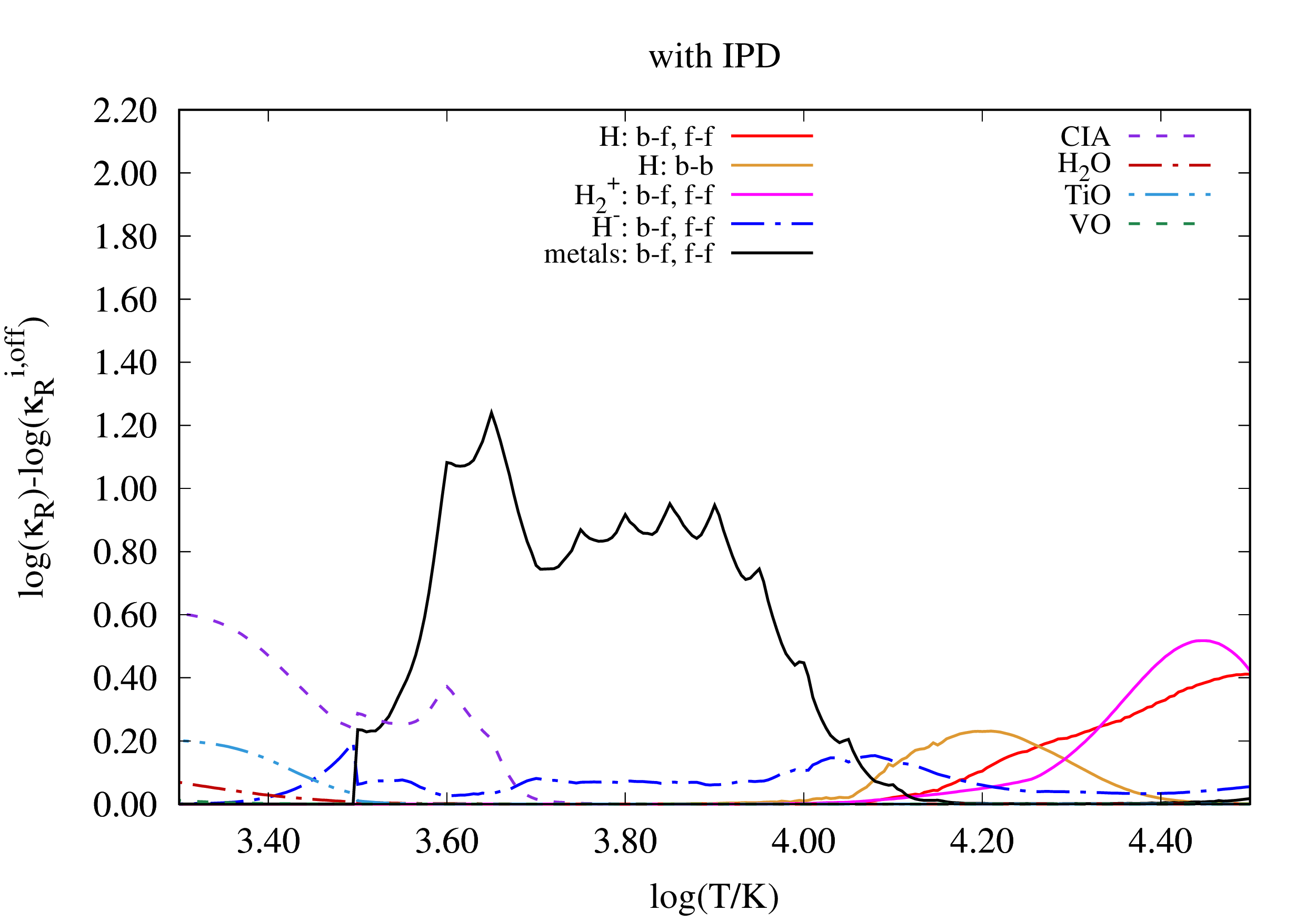}
\caption{Properties of the Rosseland mean opacity towards the highest temperature of our tables, at high density for $\log(R)=6$. The chemical composition is defined by $X = 0.7, Z = 0.01$, with a scaled-solar chemical mixture following \cite{Magg_etal_22}. Each curve corresponds to $\log(\kR)-\log(\kR^{i, {\rm off}})$, where $\kR$ is the full opacity including all opacity sources considered here, and $\kR^{i, {\rm off}}$ denotes the reduced opacity obtained by excluding the specific absorbing species. This displays the temperature window to which a given opacity source contributes the most.
The IPD effect is ignored in the computations of the left panel, whereas it is incorporated in the results of the right panel. In both cases, we apply thermal Doppler plus micro-turbulent velocity profiles for molecular transitions. }
    \label{fig_khighro}
\end{figure}

 \begin{figure}[h!]
    \centering    \includegraphics[width=0.5\textwidth]{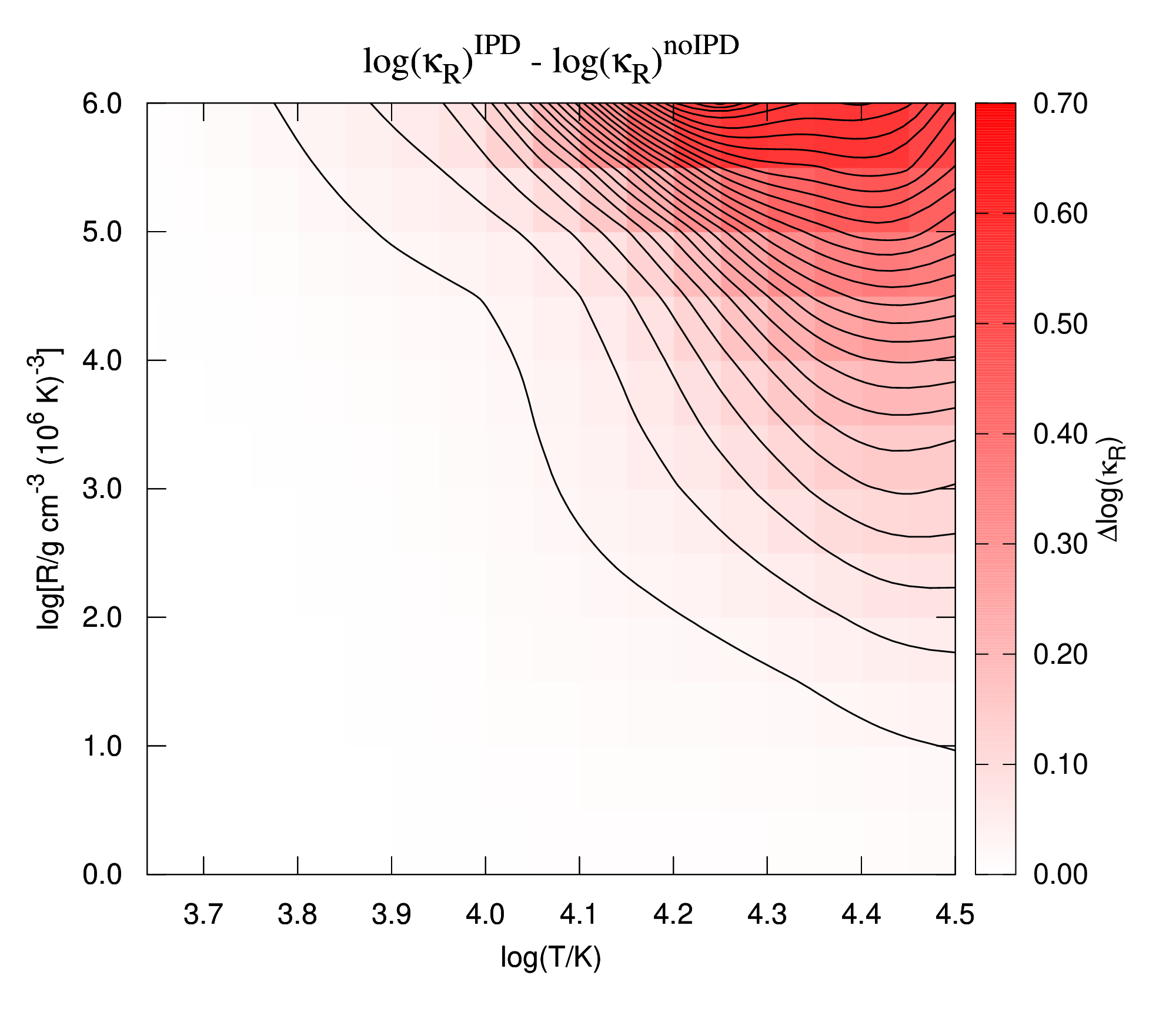}    
\caption{Map of the differences in Rosseland mean opacities when the IPD effect is included or ignored at high density. Contour levels are distributed every $0.025$ dex in $\Delta\log(\kR)$.}
    \label{fig_table_ipd}
\end{figure}

To illustrate the impact of the IPD on the opacities, we compare in Figure~\ref{fig_khighro} the different contributions to the opacity for a scaled-solar mixture, assuming $\log(R)=6$.
Moving up in temperature, we notice the significant contribution of collision-induced absorption (CIA) at low temperatures, which is primarily caused mainly by H$_2$-H$_2$ collisions.
The inclusion of the IPD for ions does not produce discernible effects in $\log(\kR)$, with differences $< 0.02$ dex up to $\log(T)=3.75$. Beyond this temperature, the contribution of different opacity sources can vary significantly as the temperature and density increase (we assume $\log(R) = 6$).
When IPD is considered, the most striking facts are: a significant decrease in H$^{-}$ opacity as the number density of neutral hydrogen decreases (see right panel of Figure~\ref{fig_potlow}), a shift to lower temperatures of the bound-bound hydrogen line opacity, and a remarkable increase in H$_2^ +$ opacity given its higher abundance.
We also notice that at high density, metals contribute significantly to opacity in the temperature range $3.5 \lesssim \log(T/{\rm K}) \lesssim 4.0$. There is a noticeable opacity bump at $3.6 \lesssim \log(T/{\rm K}) \lesssim 3.7$, and we verify that the major absorption contributions come from Fe, Al, Na, and Ca. Such a bump will be visible in the Rosseland mean opacity as well (see Section~\ref{ssec_pressureb}).

Finally, Figure~\ref{fig_table_ipd} shows a map of the differences in $\log\kR$ when the IPD effect is taken into account or neglected.
The major consequence of IPD is of increasing H and H$_2^{+}$ ionization which results in higher Rosseland mean opacity for $\log(T)> 4.1$ and $\log(R) > 1$ (red area).

\subsection{Pressure broadening effects on mean gas opacity}
\label{ssec_pressureb}
\begin{figure}[h!]
    \centering    \includegraphics[width=0.48\textwidth]{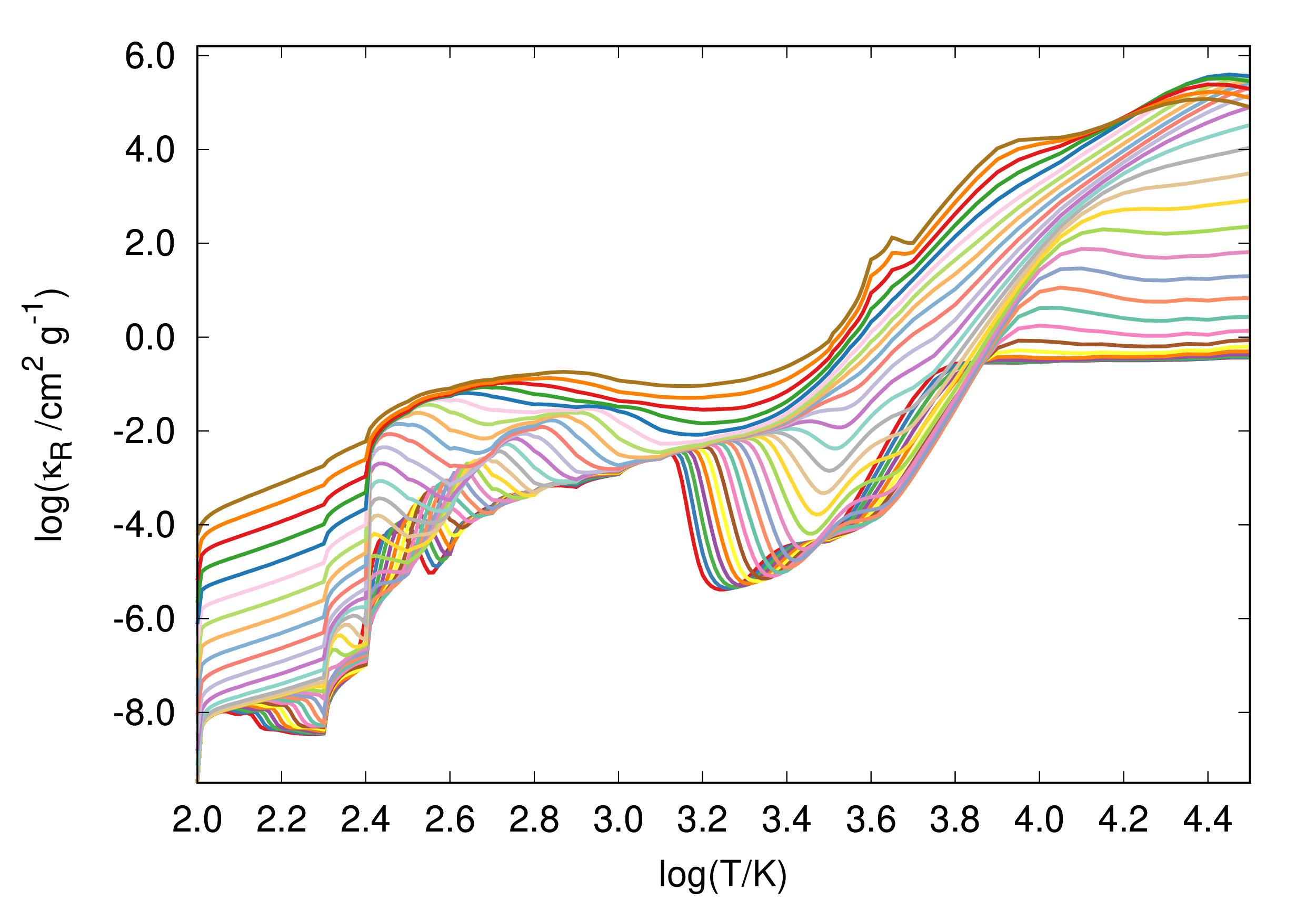}    \includegraphics[width=0.48\textwidth]{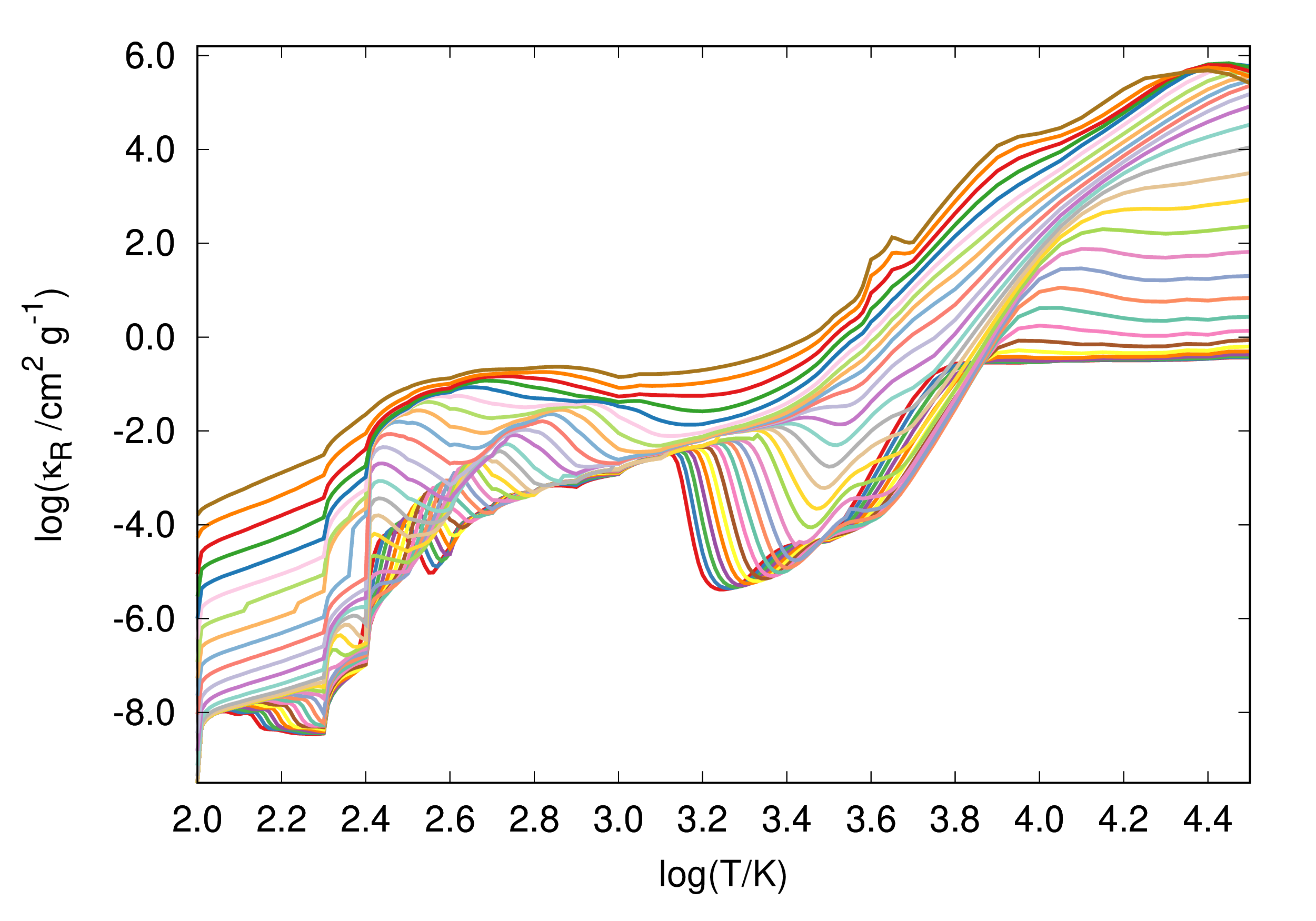}
\caption{Rosseland mean gas opacities computed in the temperature range $100 \le T/{\rm K} \lesssim 30000$ and encompassing the $R$ interval $-8 \leq \log(R) \leq 6$ in steps of 0.5 dex. The chemical composition is defined by $X = 0.7, Z = 0.01$, with a scaled-solar chemical mixture following \cite{Magg_etal_22}.
Left panel: opacities assuming Doppler plus micro-turbulent velocity molecular line profiles and ignoring the IPD effect.
Right panel: opacities assuming pressure broadening for molecular line profiles and accounting for the IPD effect.
    }
    \label{fig_kgas}
\end{figure}

\begin{figure}[h!]
    \centering    
    \centering    \includegraphics[width=0.48\textwidth]{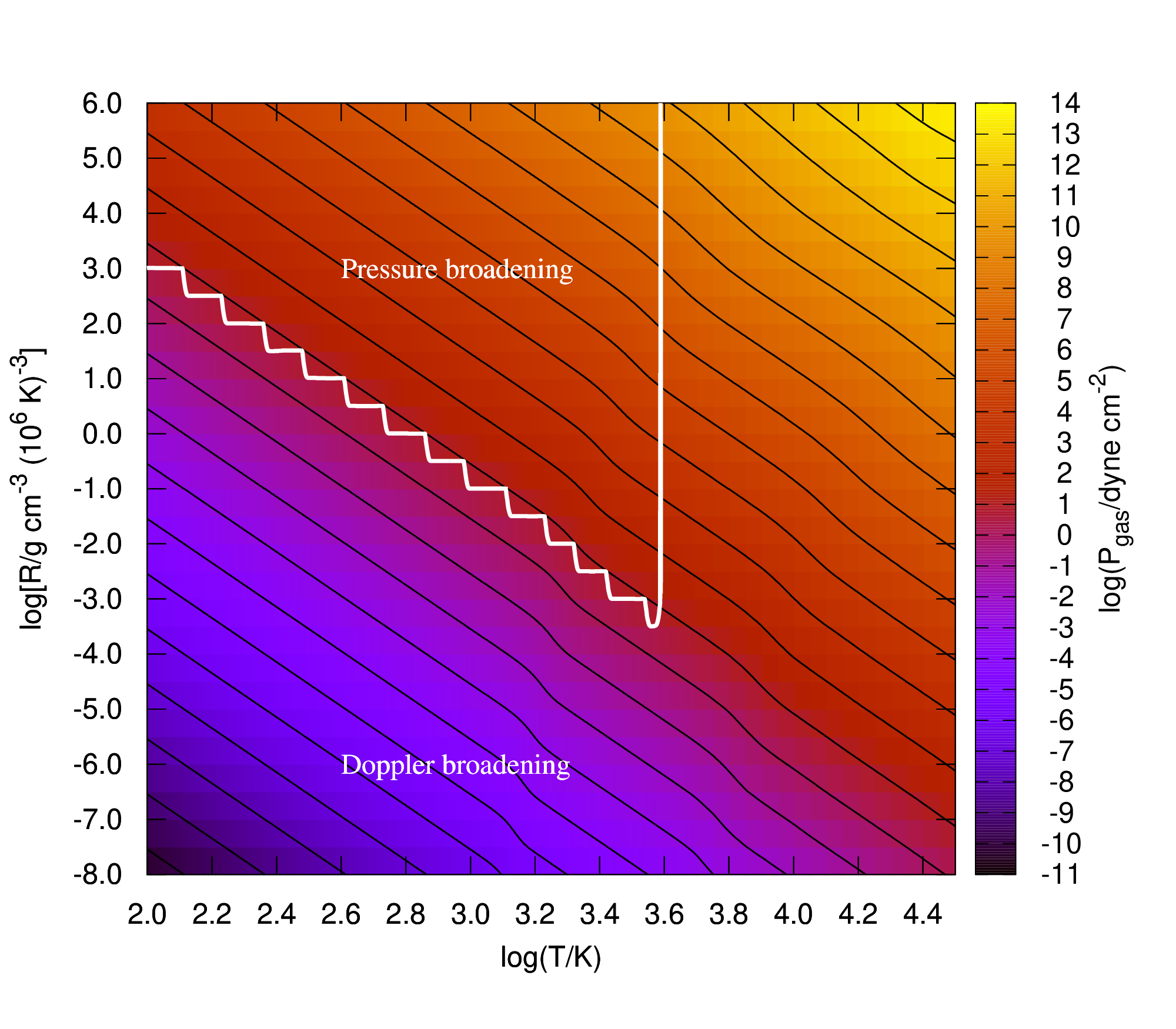} 
    \includegraphics[width=0.48\textwidth]{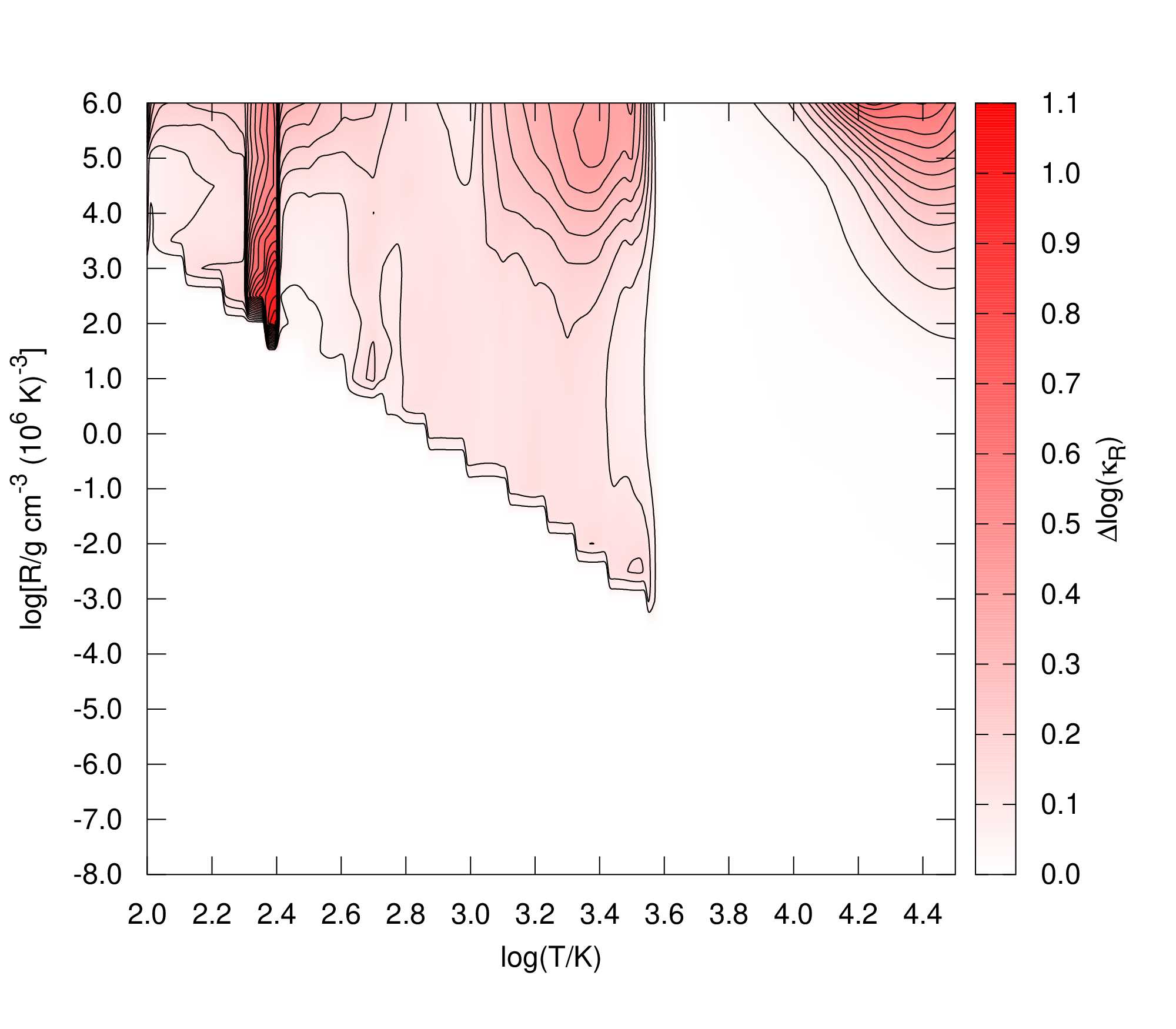}    
    \caption{
    Left panel: map of the gas pressure as a function of $T$ and $R$. Contour levels are distributed every $1$ dex in $\log(P_{\rm gas})$. The white contour draws the locus of points where the logarithmic difference in Rosseland mean opacity treated with a thermal Doppler profile and pressure broadening equals 0.001 dex in the regime where molecular transitions are important, for $T \le 4000$ K.
    Right panel: map of the differences in Rosseland mean opacities between computations that include the IPD effect and pressure broadening for molecular transitions and those that ignore the IPD and assume Doppler broadening regardless of pressure. Contour levels are distributed every $0.05$ dex in $\Delta\log(\kR)$.
    The chemical composition is the same as in Figure~\ref{fig_kgas}.}
    \label{fig_table_broad}
\end{figure}

As thoroughly discussed in several studies \citep[e.g.,][]{Freedman_etal_08,  Helling_Lucas_09, Malygin_etal_14}
mean gas opacities without a dust continuum contribution have several astrophysical applications.
They are important, for example, in a dust-depleted low-metallicity medium or when the equilibrium temperature exceeds the local dust sublimation temperature.  As a result, gas opacities can be relevant in describing the inner regions of accretion disks \citep{Muzerolle_etal_04}, calculating the energy balance of Type Ia supernovae \citep{Dessart_etal_14}, estimating the cooling of non-accreting hot white dwarfs \citep{Rohrmann_etal_12}, quantifying stellar feedback processes in the interstellar medium \citep{Pelupessy_Papadopoulos_09}, and simulating star and planet formation \citep{Helling_Lucas_09}.

To assess the impact of pressure broadening  on Rosseland mean gas opacities we performed  two independent runs of  \texttt{\AE SOPUS}, one adopting the thermal Doppler plus micro-turbulent velocity molecular line profiles, and the other assuming the Lorentzian molecular line profiles that depend on both temperature and pressure.

The  results are illustrated in Figure~\ref{fig_kgas}. We note that the dynamical range of \kR\ is extremely broad, spanning $\sim14$ orders of magnitude, making eye-comparison somewhat difficult.  Nonetheless, we reckon it is useful to show the opacity trends in the two cases.
To quantify the differences in \kR, we create a map that spans the whole area of the table, as shown in the right panel of Figure~\ref{fig_table_broad}.
In Section~\ref{ssec_potlow}, the IPD effect has already been discussed.
To help the discussion, in the left panel we also draw a map of the gas pressure, indicating the contour level (white line) above which we begin to consider pressure broadening of molecular transitions.

\subsubsection{The opacity significance of neutral alkalis Na and K}
The contribution of alkali atoms in atmospheric opacity of cool sub-stellar objects was initially established by studying the far red spectra of T dwarfs 
\citep{Burrows_etal_00}.
Atomic pressure-broadened  lines, especially those of Na and K, are
major opacity sources over certain spectral ranges, temperatures,
and densities \citep{Freedman_etal_08}.
With their large absorption cross sections at  near infra-red and far-red wavelengths, sodium and potassium fill  what would otherwise be a spectral region of relatively low opacity.

\begin{figure}[h!]
    \centering   
    \includegraphics[width=0.48\textwidth]{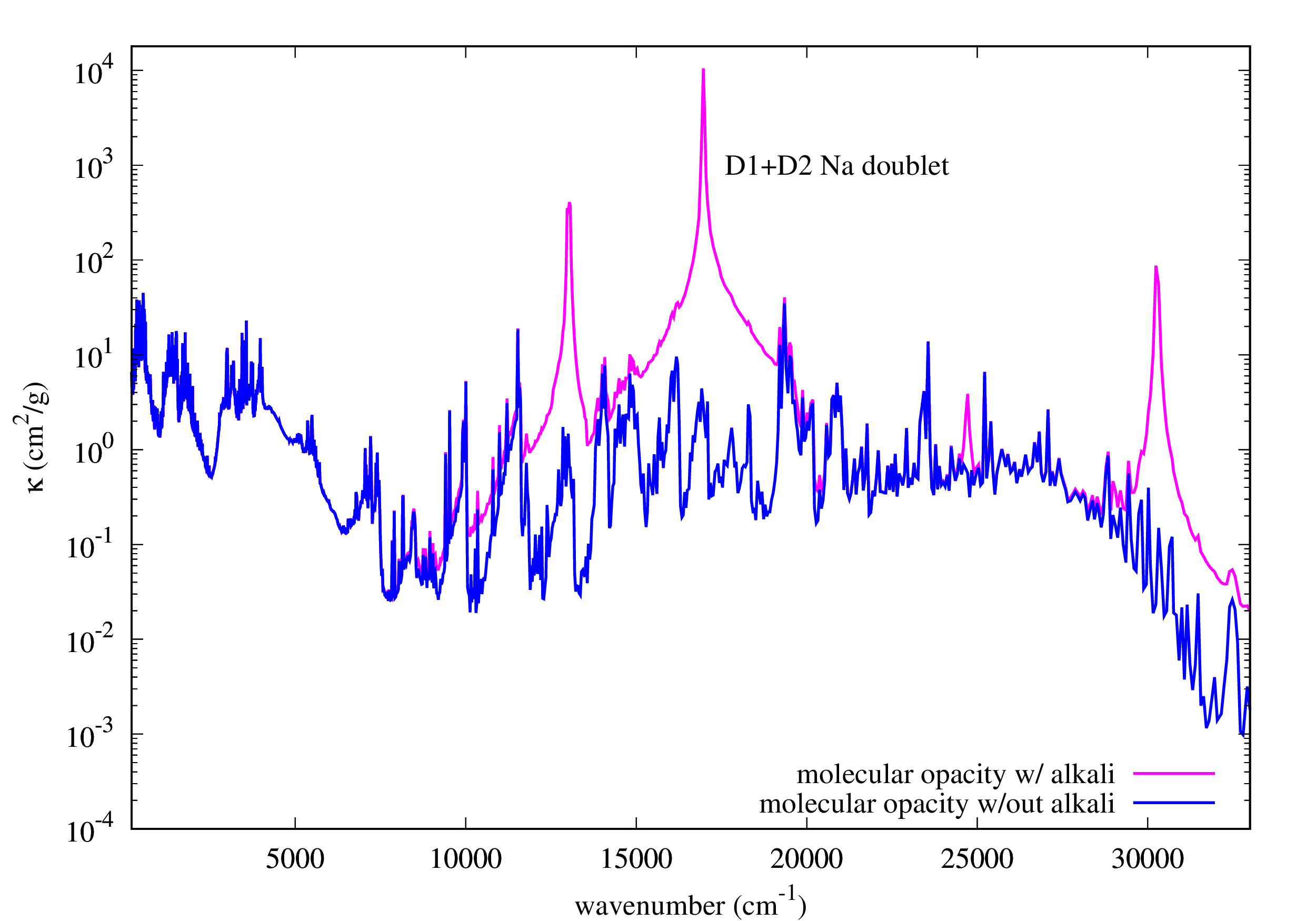}   
    \includegraphics[width=0.48\textwidth]{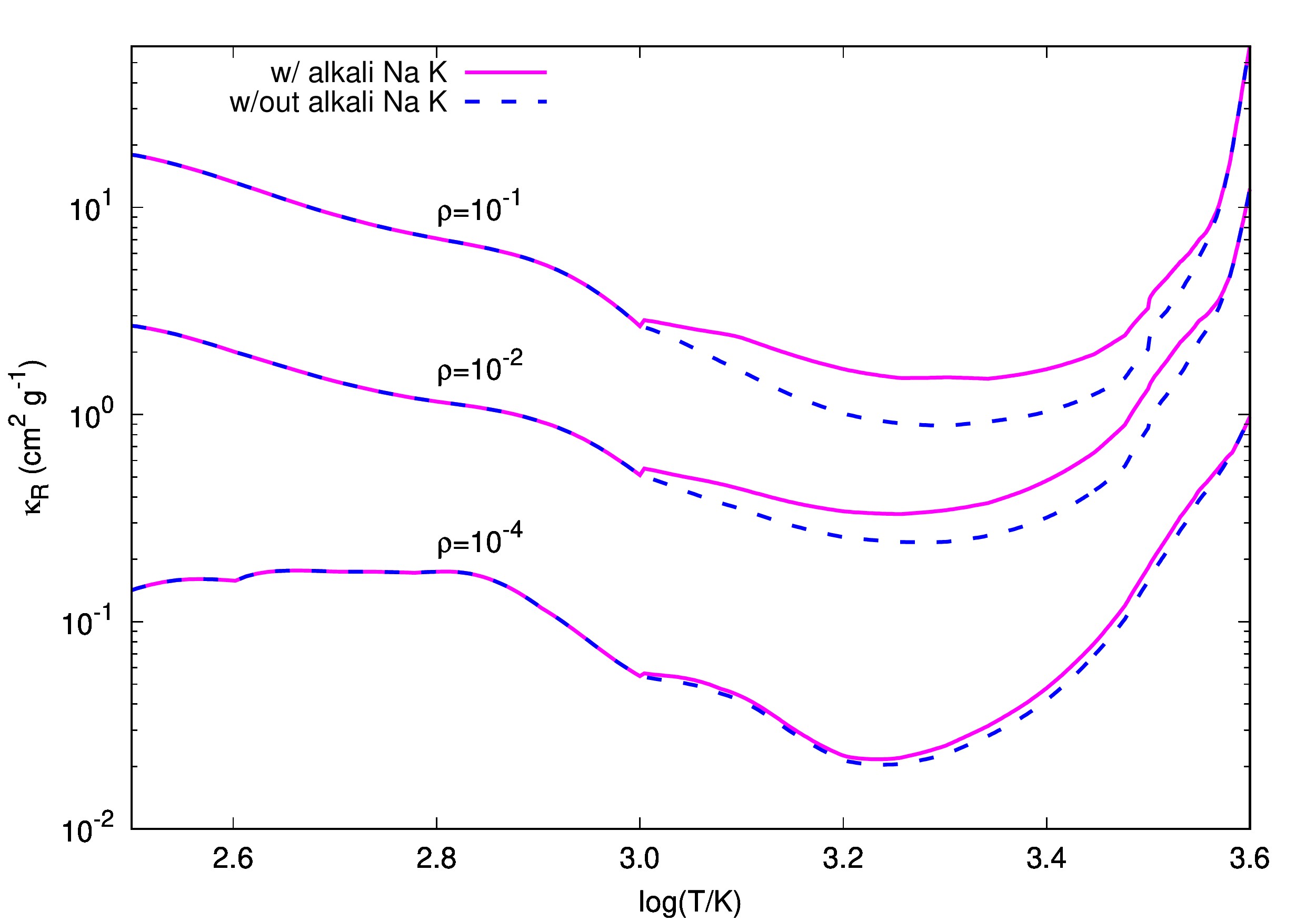}  
    \caption{Left panel: Molecular gas opacity as a function of wavenumber at $T=1585$ K, gas pressure $P_{\rm gas} =211$ bar, and density $\rho=0.004$ g ${\rm cm^{-3}}$.
    The reference solar chemical composition is taken from \cite{Magg_etal_22}, with metallicity $Z=0.02$ and hydrogen abundance $X=0.7$. The blue curve represents the monochromatic opacity without the contribution of alkali atoms Na and K, whereas the magenta curve includes the two atoms' opacity contributions. Right panel: Rosseland mean opacity as a function of temperature at three densities (g ${\rm cm^{-3}}$). Magenta lines include the opacity from alkali atoms Na and K, whereas dashed blue lines do not. The chemical composition is scaled-solar according to \cite{Magg_etal_22}, with metallicity $Z=0.01$ and hydrogen abundance $X=0.7$.}
    \label{fig_alkali}
\end{figure}
The left panel of Figure~\ref{fig_alkali} compares the total molecular opacity (magenta line) as a function of wave-number for a gas temperature of 
1585 K and a gas pressure of 211 bar,  with a calculation that does not account for alkali opacity (blue line). Above about  a wave-number of 10000 cm$^{-1}$, the alkali opacity plays a significant role in determining the total summed opacity.
The resonance Na doublet at $\simeq 17000$ cm$^{-1}$ stands out as a prominent source of absorption.
The right panel of Figure~\ref{fig_alkali}  compares the Rosseland mean opacity with and without the contribution of alkali metals at various densities. It is evident that the alkali opacity fills in the opacity minimum from about 1000 to 3200 K at higher densities.  Na and K lines play a much smaller role at lower densities, so the differences are minor.

\subsubsection{Comparison with other authors}
\begin{figure}[h!]
    \centering     \includegraphics[width=0.48\textwidth]{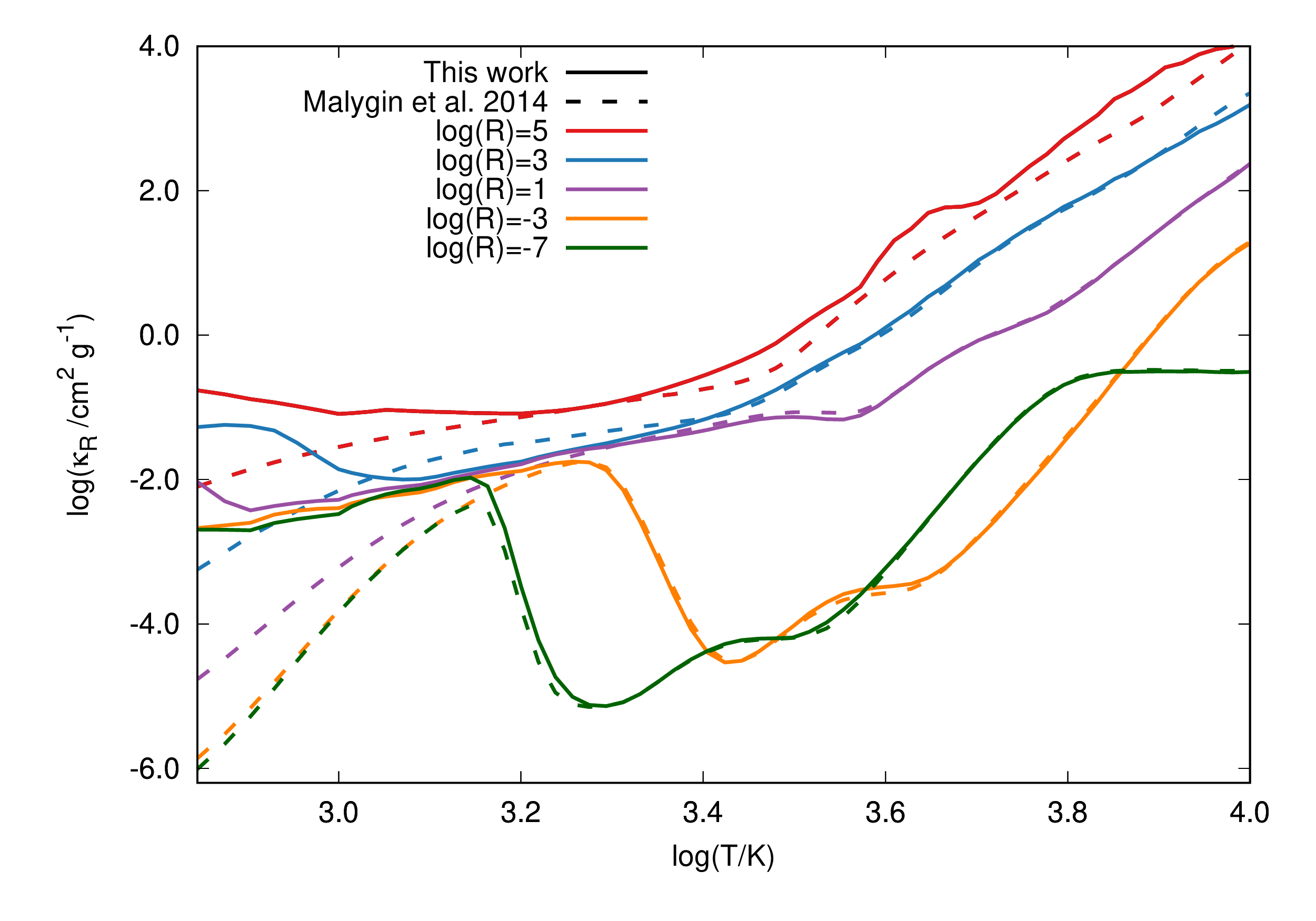}  
\includegraphics[width=0.48\textwidth]{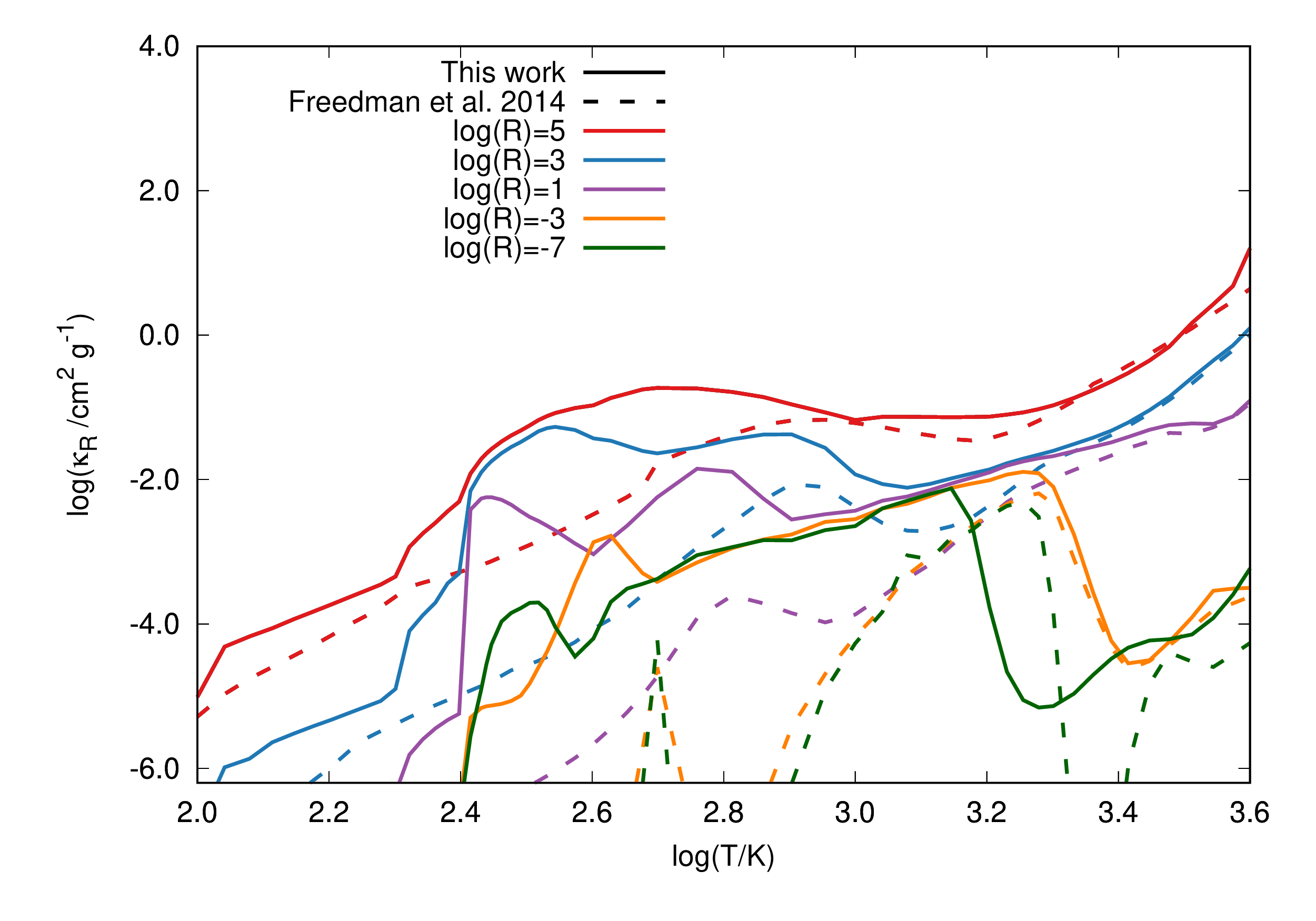}   
    \caption{Comparison of Rosseland mean gas opacities computed in this work and by other studies.
    Left panel: Comparison with \citet{Malygin_etal_14}. The reference solar chemical composition is taken from \cite{GS_98}, with metallicity $Z=0.01696$ and hydrogen abundance $X=0.7347$.
Right panel: Comparison with \citet{Freedman_etal_14}. The reference solar chemical composition is taken from \cite{Lodders_03}, with metallicity $Z=0.0133$ and hydrogen abundance $X=0.7491$.    
    Values for $\log(R)$ are labeled.}
    \label{fig_Maligyn14_Freedman14}
\end{figure}
To test our results we consider the Rosseland mean gas opacities computed by \cite{Malygin_etal_14} and \citet{Freedman_etal_14}.
In \cite{Malygin_etal_14} work Kurucz's CD-ROMs were used to extract the line and continuum opacity data \citep{Kurucz_93}, and the mean opacities were calculated using the publicly available \texttt{DFSYNTHE} code \citep{Castelli_05}.
The results are presented in the left panel of Figure~\ref{fig_Maligyn14_Freedman14} for a temperature range of 700 K to 10000 K.
The agreement between the two opacity sets is good for $\log(T)>3.6$. At lower temperatures, some deviations start to appear, most likely due to different line lists for atoms and molecules, as well as different line pressure broadening treatments. The most prominent discrepancy shows up at $\log(T)< 3.3$. While our opacity results remain either relatively flat or even increase for the highest $R$ values, \cite{Malygin_etal_14} opacities do, in fact, significantly decrease at lower temperatures.
For example, our opacity curve for $\log(R)=-7$ only includes thermal Doppler profiles for the molecular transitions, as pressure is irrelevant in this case. Despite this, the decrease at lower temperatures is much less pronounced than in \cite{Malygin_etal_14}.
One plausible explanation is that we rely on molecular line lists that extend down to 100 K in temperature, whereas in Kurucz opacity the contributions of molecules cease for $T<1995$ K, and therefore extrapolations below that limit  may be inaccurate.

In comparison to \citet{Freedman_etal_14} results shown in the right panel of Figure~\ref{fig_Maligyn14_Freedman14}, there are significant differences in Rosseland mean opacities. We found no explicit information about pressure broadening for molecular transitions, and molecular absorption is limited to 12 species. Conversely, we consider 80 absorbing molecules in our study.
Both facts could explain the disparity in outcomes.

\section{Rosseland mean opacities with solid grains}
\label{sec_solidgrains}
We present a few examples of Rosseland mean opacities with solid grains included. The dust prescriptions are the same as in \citet{marigo23}.
We are aware that dust clouds in brown dwarfs and planets are crucial for understanding their atmospheric properties, formation mechanisms, and overall behavior, providing valuable insights into the broader field of planetary science. Dust clouds can significantly impact their atmospheric composition. They often consist of various particles, including silicates, iron, and other compounds, contributing to the chemical makeup of the atmosphere. In this context, important contributions were provided by \citet{Tsuji_etal_96}, \citet{Burrows_etal_00}, \citet{Ackerman_Marley_01}, \citet{Tsuji_etal_02}, \citet{Sharp_Burrows_07},  \citet{Helling_etal_08}, \citet{Witte_etal_09}, \citet{Allard_etal_12}, \citet{Juncher_etal_17}, \citet{Woitke_etal_20}.
Because our primary interest is in very low mass stars, in this work we do not take into account the formation of dust clouds and we postpone the effort to a future study.

\begin{figure}
\centering
\includegraphics[width=0.48\textwidth]{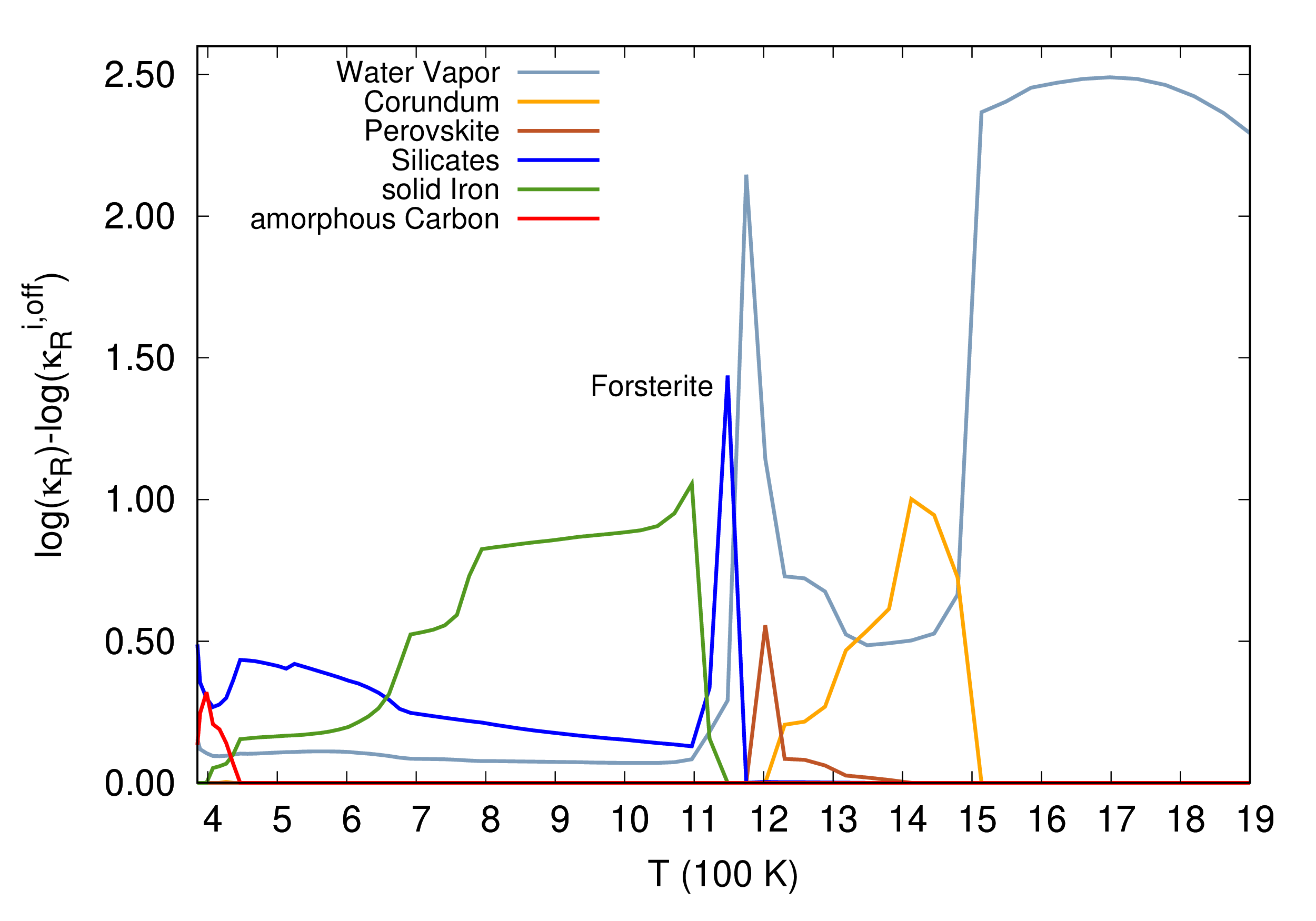}
\includegraphics[width=0.48\textwidth]{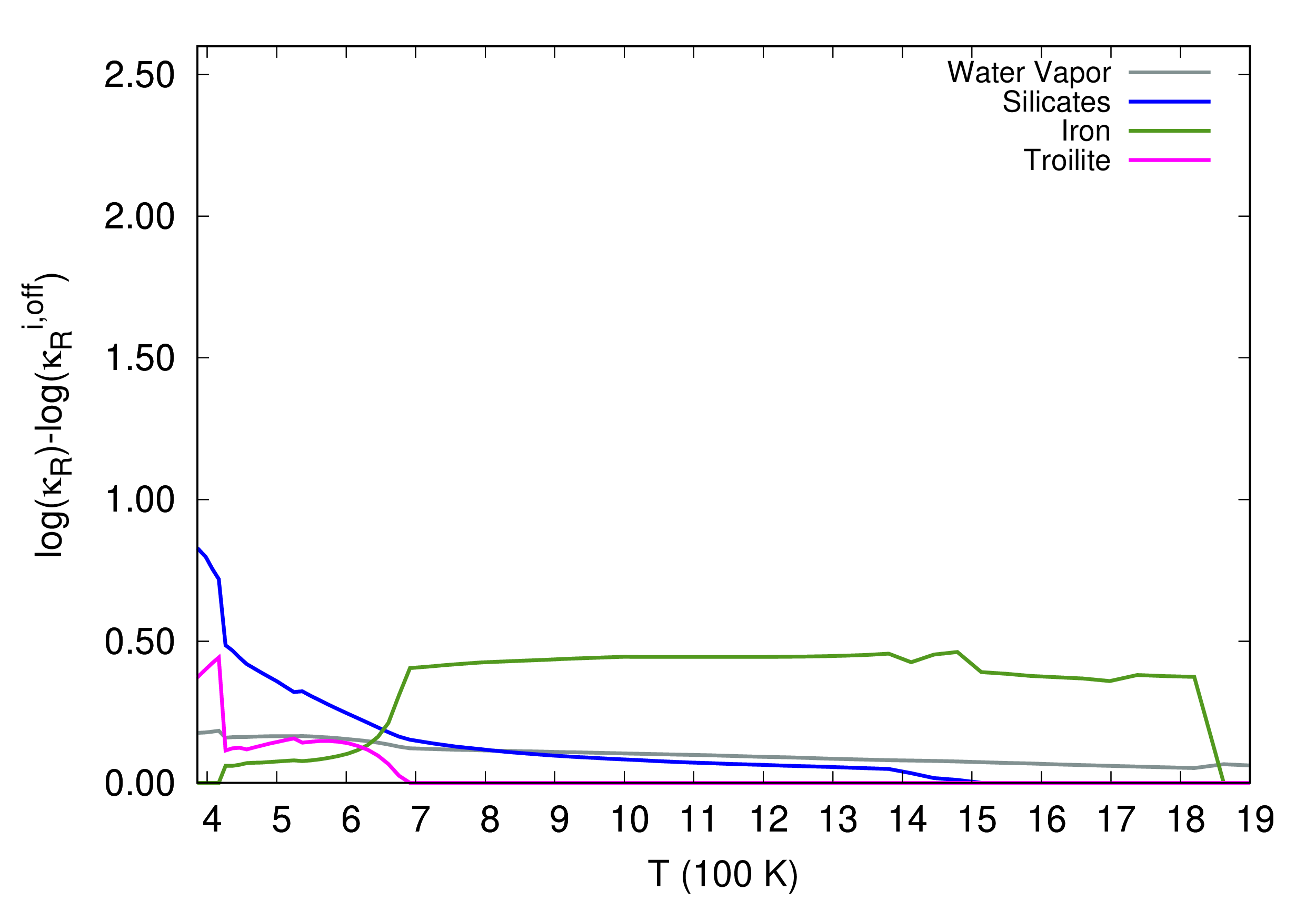}
    \caption{Properties of the Rosseland mean opacity for temperatures where solid grains dominate,
    at low density ($\log(R)=-3$, left panel) and high density ($\log(R)=6$, right panel).    
    The chemical composition is defined by $X = 0.735,\, Z = 0.0165$, with scaled-solar elemental abundances following \cite{Magg_etal_22}. 
      Each curve represents the contribution of major solid species to the total Rosseland mean opacity, and it is calculated as  $\log(\kR)-\log(\kR^{i, {\rm off}})$, where $\kR$ is the full opacity including all opacity sources considered here, and $\kR^{i, {\rm off}}$ is the reduced opacity computed by excluding the specific absorbing species.}
    \label{fig_kgrains}
\end{figure}

Figure~\ref{fig_kgrains} compares 
major solid species at low and high densities that contribute most to the opacity. There are  significant differences between the two regimes.

First, we notice that the corundum opacity bump that appears for $\log(R)=-3$ at temperatures ranging from 1500 to 1200 K is missing for $\log(R)=6$. Al$_2$O$_3$ does not condense at high pressure conditions.
Even at $\log(R)=-3$, where corundum contributes to opacity, molecular band absorption by water molecules continues to play a significant role in opacity. This extends down to around 400 K.
In contrast, for $\log(R)=6$, the opacity contribution of water vapor is significantly reduced.
Another major distinction is that at high density, solid iron is the dominant opacity source from $\simeq 1870$ K to 650 K, whereas at lower temperatures, silicates begin to prevail.
Furthermore, we see that amorphous carbon does not condense, whereas Troilite (FeS) has a discernible contribution, from about 700 K to 380 K.
\begin{figure}
    \centering    \includegraphics[width=0.60\textwidth]{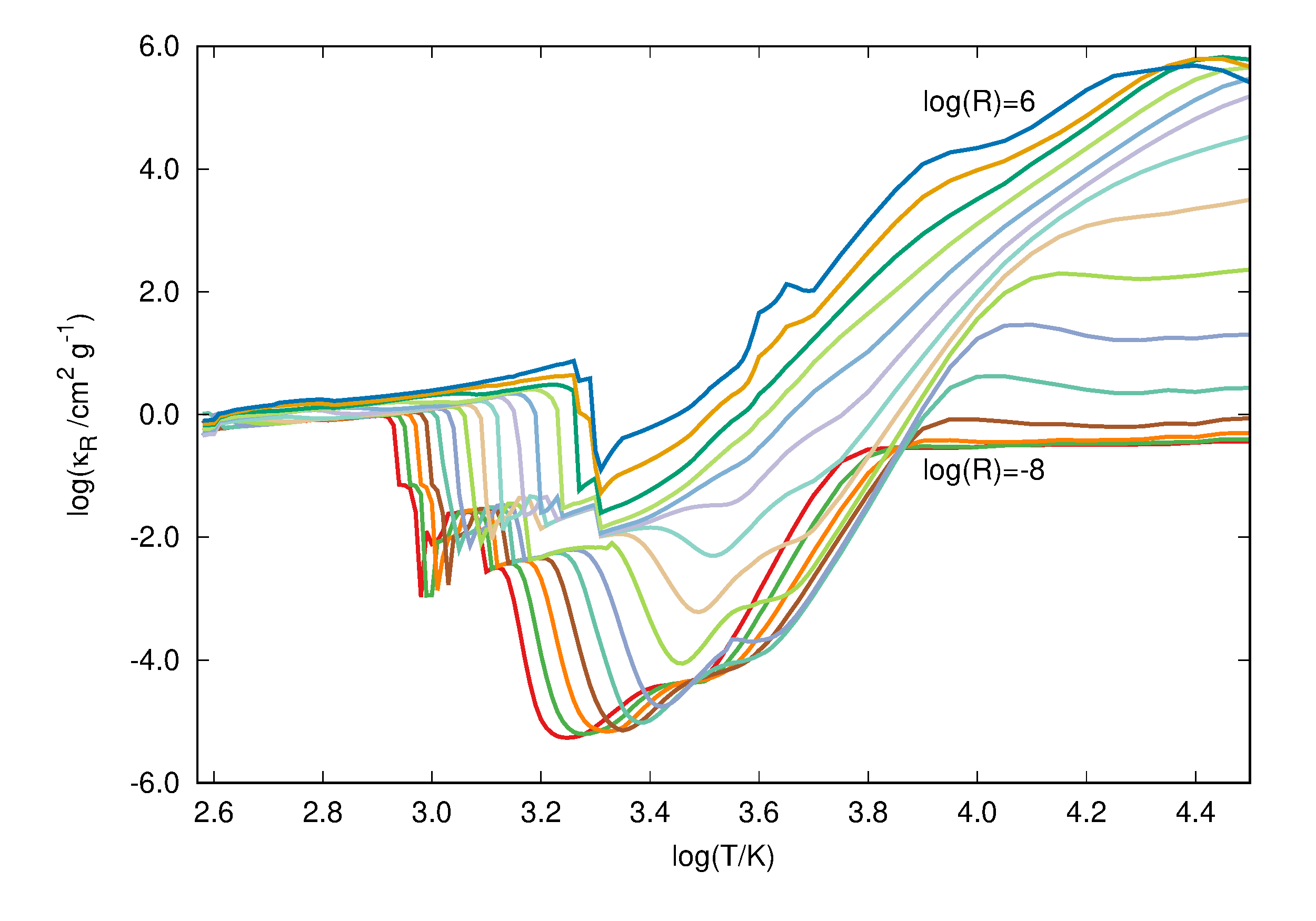}
    \caption{Rosseland mean opacities with the inclusion of condensed dust grains. The chemical composition is the same as in Figure~\ref{fig_kgrains}. The curves are distributed every 1 dex in $\log(R)$, in the interval $-8\le\log(R)\le 6$.}
    \label{fig_grains}
\end{figure}

Figure~\ref{fig_grains} shows the behavior of Rosseland mean opacities over the range $-8\le\log(R)\le 6$. As previously discussed, we see that the corundum opacity bump (at $3.08\lesssim\log(T)\lesssim3.17$) is present for $\log(R)< 2$, while for $\log(R)> 2$ the species does not condense and solid iron makes the most important opacity contribution for $650 \lesssim T/{\rm K}\lesssim 1870$ ($2.81\lesssim\log(T)\lesssim3.27$).

\section{Impact on stellar models}
\label{sec_impactmodels}
\begin{figure}[h!]
    \centering     \includegraphics[width=\textwidth]{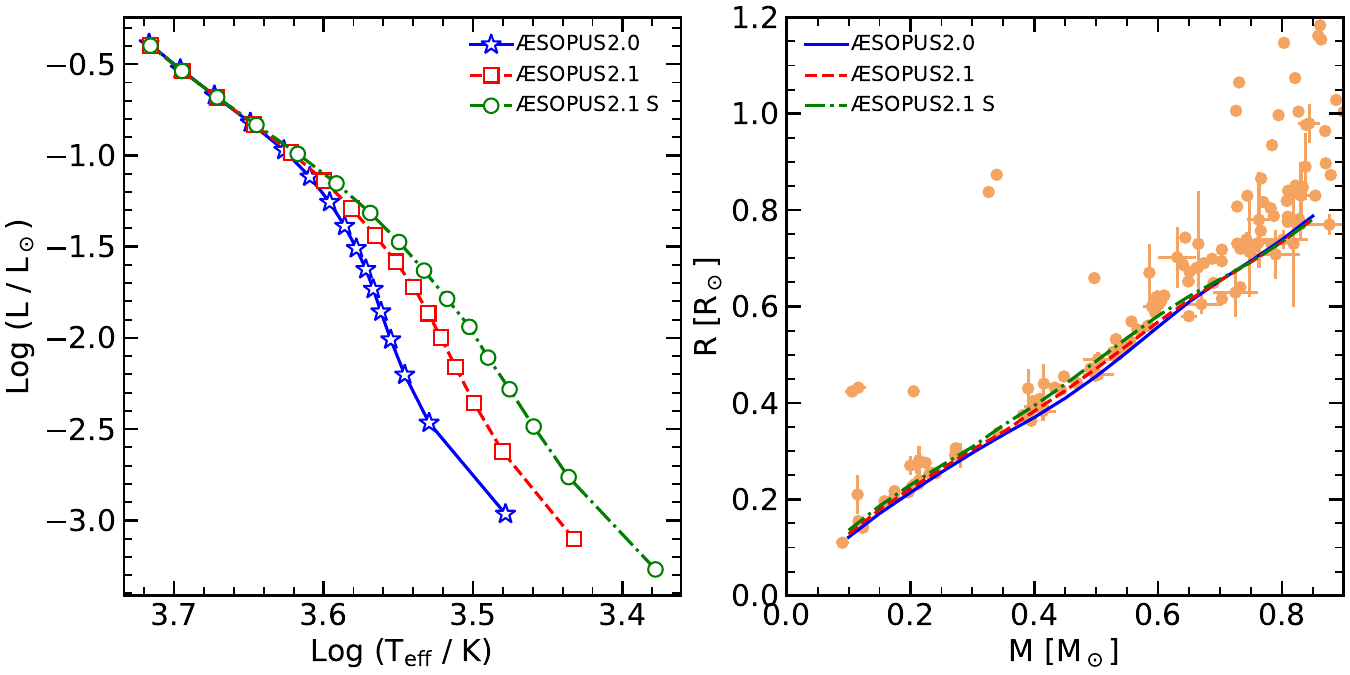}  
    \caption{The impact of our new Rosseland mean opacities on low-mass stellar models computed with the PARSEC v2.0 code. The left panel shows the HR diagram for models in the mass range from 0.1 to 0.85 \Msun\ and computed at intervals of 0.05 \Msun\ (from bottom-right to top-left), and plotted at ages of 5 Gyr. The blue symbols indicate models computed with the previous version of the opacity tables from \citet{aesopus2}, while the red and green symbols are for models computed with present opacity tables. The green symbols correspond to models with a shift in their $T-\tau$ relation (see text for details). The right panel shows the same models in the mass--radius plane, now compared with the empirical data of low-mass stars in double-lined eclipsing binary catalogs from \citet{segransan03, demory09, torres10, carter11, doyle11, kraus11, parsons12a, parsons12b, debcat}. }
    \label{fig_newmodels}
\end{figure}

\begin{figure}[h!]
    \centering     \includegraphics[width=\textwidth]{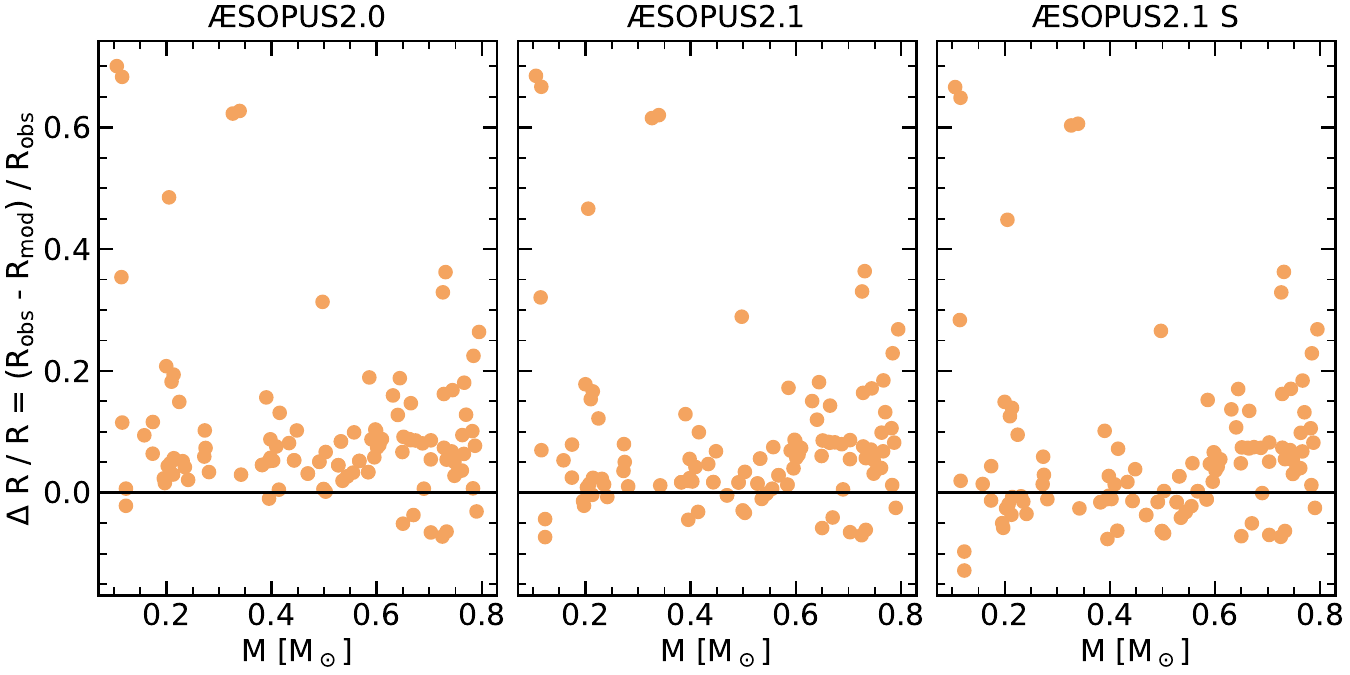}  
    \caption{The relative difference between the observed and model radii for the 3 sets of models and the data presented in the right panel of Fig.~\ref{fig_newmodels}, as a function of stellar mass. It can be seen that while our new opacities reduce the systematic discrepancies between the predicted and observed radii for stars in the $M<0.6$~\Msun\ range, a shift in the $T-\tau$ relation (or other alternatives as discussed in the text) seems to be still necessary to eliminate it completely.}
    \label{fig_newmodels_residuals}
\end{figure}

Figure~\ref{fig_vlm} suggests that the updated Rosseland mean opacities could have a significant impact on stellar models of masses below $\sim0.7$~\Msun, particularly influencing the temperature gradient in their superadiabatic regions. This conjecture holds true, as illustrated in Fig.~\ref{fig_newmodels}, depicting the consequences of employing the revised opacities on the fundamental properties of \texttt{PARSEC} models within the mass range of 0.1 to 0.85 \Msun. These results are compared with models utilizing the prior version of \texttt{\AE SOPUS} opacities \citep[specifically, \texttt{\AE SOPUS} v2.0 from][]{aesopus2}. In this latter case, the opacities were not readily available for $\log(R)$ values exceeding 1, prompting PARSEC to use the opacity values taken at the border of the available tables. This assumption was grounded in the expectation that, in the extensive convective regions of such stars, the temperature gradient would approach adiabatic conditions, rendering it largely insensitive to variations in the opacity tables.

The models in Figure~\ref{fig_newmodels} are computed with \texttt{PARSEC} code version 2.0 \citep{costa19, nguyen22}, and adopting $T-\tau$ relations interpolated from the \texttt{PHOENIX} stellar atmosphere models \citep{Allard_etal_12}, implemented and discussed in \citet{chen14}. Here, suffice it to recall that we use  the solar composition from \citet{Caffau_etal_2011}\footnote{In this case, the \citet{Caffau_etal_2011} composition is used in \texttt{PARSEC} and in the opacity tables, but not in the \texttt{PHOENIX} atmosphere models, which are based on the \citet{Asplund_etal_09} solar composition. Obtaining full consistency between all these components is beyond our reach at the moment.} and the mixing length theory with a parameter $\alpha=1.74$ derived from the calibration of the Solar model \citep{Bressan_etal_12}. We present models either using (labelled with S) or not using the shift in the $T-\tau$ relation advocated by \citet{chen14}. Further details will be discussed in a subsequent paper dedicated to very-low mass star models. 

Since low-mass models evolve minimally after settling into their main sequences, only their properties at the age of 5 Gyr are presented in Figure~\ref{fig_newmodels}. From the Hertzsprung-Russell (HR) diagram, it is evident that the use of new opacities consistently produces cooler and fainter models. Furthermore, the impact of the new opacities diminishes as we approach stellar models with a mass of 0.6 $\Msun$, as expected for stellar structures that predominantly evolve in the range of $\log(R)<1$ (refer to Figure~\ref{fig_vlm}).

The right panel of Figure~\ref{fig_newmodels} displays the same models in the mass-radius plane. In this instance, the empirical data derived from double-lined eclipsing binary catalogs is superimposed. As observed, our PARSEC models computed with \texttt{\AE SOPUS} v2.0 opacities generally align with the lower limit of the empirical mass-radius relation. This once again highlights the ongoing manifestation of the mass-radius discrepancy, extensively documented in various studies \citep[see, for example,][]{chen14,torres14,somers20}. It is evident that the use of the new \texttt{\AE SOPUS} v2.1  opacities somewhat mitigates this discrepancy by slightly inflating all models in the 0.1-0.6\ \Msun\ range.
Therefore, we advocate the use of proper opacity tables extended to high densities and pressures, to describe very-low mass models. 

Figure~\ref{fig_newmodels_residuals} shows the relative differences between the observed and model radii, for the same models and data as in Fig.~\ref{fig_newmodels}, so as to allow a better visualisation of the small improvements reached by using  \texttt{\AE SOPUS} v2.1 instead of v2.0 opacities. An additional panel presented additional models, labelled \texttt{\AE SOPUS} 2.1 S, in which the $T-\tau$ relation is shifted by just half the amount calibrated in \citet{chen14} -- that is, these models adopt a shift of $\Delta\log(T/\Teff)=0.03$ dex for $\log(\Teff/\mathrm{K}) < 3.5$, and gradually reduce this shift to 0 for $\log(\Teff/\mathrm{K})$ between 
3.5 and 3.765. These latter models practically cancel out the discrepancy in the mass-radius relation at low masses. 

Therefore, while our new \texttt{\AE SOPUS} v2.1 opacities reduce the systematic discrepancies between the predicted and observed radii of very low mass stars, additional model assumptions -- like the shift in the $T-\tau$ relation by \citet{chen14} or the stellar spots by \citet[][]{somers20} -- seem to be still necessary to eliminate them completely.

\section{Concluding Remarks}
\label{sec_concl}
We compute the equation of state and provide Rosseland mean opacity tables for temperatures ranging from 32,000 K to 100 K. These tables are expected to be useful for a series of applications, which we leave to the readers to explore. Tables for different chemical compositions can be retrieved with the updated \texttt{\AE SOPUS} web interface in \url{http://stev.oapd.inaf.it/aesopus}\footnote{The long-lasting link for \texttt{\AE SOPUS 2.1} web interface is \url{http://stev.oapd.inaf.it/cgi-bin/aesopus_2.1}. We recall that the previous versions of \texttt{\AE SOPUS} are still available in their original locations at \url{http://stev.oapd.inaf.it/cgi-bin/aesopus_1.0} and \url{http://stev.oapd.inaf.it/cgi-bin/aesopus_2.0}.}, where one will also find links to a set of pre-computed tables for 10 popular chemical mixtures (i.e.\citealt{AG_1989}, \citealt{GN_93}, \citealt{GS_98}, \citealt{Holweger_01}, \citealt{Lodders_03}, \citealt{Grevesse_etal_07}, \citealt{Asplund_etal_09}, \citealt{Caffau_etal_2011}, \citealt{Asplund_etal_21}, and \citealt{Magg_etal_22}), and in each case spanning wide ranges in Z and X values. We have so far verified that these new opacity tables lead to important changes in the modeling of very low mass stars at near-solar metallicities, causing shifts in their mass--effective temperature and mass-luminosity relations. More modest shifts are present in the mass-radius relation, and they go in the right direction to alleviate the discrepancies in the radii of very low mass stars that have been widely reported in the literature.

\begin{acknowledgments}
We acknowledge support from Padova University through the research project PRD 2021. PM and AB acknowledge the Italian Ministerial grant PRIN2022, ``Radiative opacities for astrophysical applications'', no. 2022NEXMP8. GC acknowledges support from the Agence Nationale de la Recherche grant POPSYCLE number ANR-19-CE31-0022. LG acknowledges partial support from an INAF Theory Grant 2022. 
\end{acknowledgments}

\software{\texttt{\AE SOPUS} \citep{aesopus2, Marigo_Aringer_09},  
    \texttt{EXOCROSS} \citep{EXOCROSS_2018}, \texttt{DIANA} \citep{Woitke_etal_16},
    \texttt{GGchem} \citep{ggchem_18}
          }

\newpage
\bibliography{opachighrho.bib}{}
\bibliographystyle{aasjournal}

\appendix
\section{Chemical species considered in \texttt{\AE SOPUS}}
\label{sec_appendix}

For completeness, Table~\ref{tab:polyatomic} presents the complete list of chemical species considered at the moment in the equation of state of \texttt{\AE SOPUS 2.1}. 
\begin{table}[ht]
\caption{Chemical species considered in the EOS of \texttt{\AE SOPUS}.}
\centering
\scalebox{0.75}{%
\begin{tabular}{l l l l l l l l l l l l l}
\hline
$\mathrm{H   ^{} }$            & $\mathrm{H   ^{+} }$           & $\mathrm{He   ^{} }$           & $\mathrm{He   ^{+} }$          & $\mathrm{He   ^{++} }$         & $\mathrm{Li   ^{} }$           & $\mathrm{Li   ^{+} }$          & $\mathrm{Li   ^{++} }$         & $\mathrm{Be   ^{} }$           & $\mathrm{Be   ^{+} }$          & $\mathrm{Be   ^{++} }$         & $\mathrm{B   ^{} }$            & $\mathrm{B   ^{+} }$           \\
$\mathrm{B   ^{++} }$          & $\mathrm{C   ^{} }$            & $\mathrm{C   ^{+} }$           & $\mathrm{C   ^{++} }$          & $\mathrm{C   ^{+++} }$         & $\mathrm{C   ^{++++} }$        & $\mathrm{N   ^{} }$            & $\mathrm{N   ^{+} }$           & $\mathrm{N   ^{++} }$          & $\mathrm{N   ^{+++} }$         & $\mathrm{N   ^{++++} }$        & $\mathrm{O   ^{} }$            & $\mathrm{O   ^{+} }$           \\
$\mathrm{O   ^{++} }$          & $\mathrm{O   ^{+++} }$         & $\mathrm{O   ^{++++} }$        & $\mathrm{O   ^{+++++} }$       & $\mathrm{F   ^{} }$            & $\mathrm{F   ^{+} }$           & $\mathrm{F   ^{++} }$          & $\mathrm{F   ^{+++} }$         & $\mathrm{F   ^{++++} }$        & $\mathrm{Ne   ^{} }$           & $\mathrm{Ne   ^{+} }$          & $\mathrm{Ne   ^{++} }$         & $\mathrm{Ne   ^{+++} }$        \\
$\mathrm{Ne   ^{++++} }$       & $\mathrm{Ne   ^{+++++} }$      & $\mathrm{Na   ^{} }$           & $\mathrm{Na   ^{+} }$          & $\mathrm{Na   ^{++} }$         & $\mathrm{Na   ^{+++} }$        & $\mathrm{Na   ^{++++} }$       & $\mathrm{Mg   ^{} }$           & $\mathrm{Mg   ^{+} }$          & $\mathrm{Mg   ^{++} }$         & $\mathrm{Mg   ^{+++} }$        & $\mathrm{Mg   ^{++++} }$       & $\mathrm{Al   ^{} }$           \\
$\mathrm{Al   ^{+} }$          & $\mathrm{Al   ^{++} }$         & $\mathrm{Al   ^{+++} }$        & $\mathrm{Al   ^{++++} }$       & $\mathrm{Si   ^{} }$           & $\mathrm{Si   ^{+} }$          & $\mathrm{Si   ^{++} }$         & $\mathrm{Si   ^{+++} }$        & $\mathrm{Si   ^{++++} }$       & $\mathrm{P   ^{} }$            & $\mathrm{P   ^{+} }$           & $\mathrm{P   ^{++} }$          & $\mathrm{P   ^{+++} }$         \\
$\mathrm{P   ^{++++} }$        & $\mathrm{S   ^{} }$            & $\mathrm{S   ^{+} }$           & $\mathrm{S   ^{++} }$          & $\mathrm{S   ^{+++} }$         & $\mathrm{S   ^{++++} }$        & $\mathrm{Cl   ^{} }$           & $\mathrm{Cl   ^{+} }$          & $\mathrm{Cl   ^{++} }$         & $\mathrm{Cl   ^{+++} }$        & $\mathrm{Cl   ^{++++} }$       & $\mathrm{Ar   ^{} }$           & $\mathrm{Ar   ^{+} }$          \\
$\mathrm{Ar   ^{++} }$         & $\mathrm{Ar   ^{+++} }$        & $\mathrm{Ar   ^{++++} }$       & $\mathrm{K   ^{} }$            & $\mathrm{K   ^{+} }$           & $\mathrm{K   ^{++} }$          & $\mathrm{K   ^{+++} }$         & $\mathrm{K   ^{++++} }$        & $\mathrm{Ca   ^{} }$           & $\mathrm{Ca   ^{+} }$          & $\mathrm{Ca   ^{++} }$         & $\mathrm{Ca   ^{+++} }$        & $\mathrm{Ca   ^{++++} }$       \\
$\mathrm{Sc   ^{} }$           & $\mathrm{Sc   ^{+} }$          & $\mathrm{Sc   ^{++} }$         & $\mathrm{Sc   ^{+++} }$        & $\mathrm{Sc   ^{++++} }$       & $\mathrm{Ti   ^{} }$           & $\mathrm{Ti   ^{+} }$          & $\mathrm{Ti   ^{++} }$         & $\mathrm{Ti   ^{+++} }$        & $\mathrm{Ti   ^{++++} }$       & $\mathrm{V   ^{} }$            & $\mathrm{V   ^{+} }$           & $\mathrm{V   ^{++} }$          \\
$\mathrm{V   ^{+++} }$         & $\mathrm{V   ^{++++} }$        & $\mathrm{Cr   ^{} }$           & $\mathrm{Cr   ^{+} }$          & $\mathrm{Cr   ^{++} }$         & $\mathrm{Cr   ^{+++} }$        & $\mathrm{Cr   ^{++++} }$       & $\mathrm{Mn   ^{} }$           & $\mathrm{Mn   ^{+} }$          & $\mathrm{Mn   ^{++} }$         & $\mathrm{Mn   ^{+++} }$        & $\mathrm{Mn   ^{++++} }$       & $\mathrm{Fe   ^{} }$           \\
$\mathrm{Fe   ^{+} }$          & $\mathrm{Fe   ^{++} }$         & $\mathrm{Fe   ^{+++} }$        & $\mathrm{Fe   ^{++++} }$       & $\mathrm{Co   ^{} }$           & $\mathrm{Co   ^{+} }$          & $\mathrm{Co   ^{++} }$         & $\mathrm{Co   ^{+++} }$        & $\mathrm{Co   ^{++++} }$       & $\mathrm{Ni   ^{} }$           & $\mathrm{Ni   ^{+} }$          & $\mathrm{Ni   ^{++} }$         & $\mathrm{Ni   ^{+++} }$        \\
$\mathrm{Ni   ^{++++} }$       & $\mathrm{Cu   ^{} }$           & $\mathrm{Cu   ^{+} }$          & $\mathrm{Cu   ^{++} }$         & $\mathrm{Zn   ^{} }$           & $\mathrm{Zn   ^{+} }$          & $\mathrm{Zn   ^{++} }$         & $\mathrm{Ga   ^{} }$           & $\mathrm{Ga   ^{+} }$          & $\mathrm{Ga   ^{++} }$         & $\mathrm{Ge   ^{} }$           & $\mathrm{Ge   ^{+} }$          & $\mathrm{Ge   ^{++} }$         \\
$\mathrm{As   ^{} }$           & $\mathrm{As   ^{+} }$          & $\mathrm{As   ^{++} }$         & $\mathrm{Se   ^{} }$           & $\mathrm{Se   ^{+} }$          & $\mathrm{Se   ^{++} }$         & $\mathrm{Br   ^{} }$           & $\mathrm{Br   ^{+} }$          & $\mathrm{Br   ^{++} }$         & $\mathrm{Kr   ^{} }$           & $\mathrm{Kr   ^{+} }$          & $\mathrm{Kr   ^{++} }$         & $\mathrm{Rb   ^{} }$           \\
$\mathrm{Rb   ^{+} }$          & $\mathrm{Rb   ^{++} }$         & $\mathrm{Sr   ^{} }$           & $\mathrm{Sr   ^{+} }$          & $\mathrm{Sr   ^{++} }$         & $\mathrm{Y   ^{} }$            & $\mathrm{Y   ^{+} }$           & $\mathrm{Y   ^{++} }$          & $\mathrm{Zr   ^{} }$           & $\mathrm{Zr   ^{+} }$          & $\mathrm{Zr   ^{++} }$         & $\mathrm{Nb   ^{} }$           & $\mathrm{Nb   ^{+} }$          \\
$\mathrm{Nb   ^{++} }$         & $\mathrm{Mo   ^{} }$           & $\mathrm{Mo   ^{+} }$          & $\mathrm{Mo   ^{++} }$         & $\mathrm{Tc   ^{} }$           & $\mathrm{Tc   ^{+} }$          & $\mathrm{Tc   ^{++} }$         & $\mathrm{Ru   ^{} }$           & $\mathrm{Ru   ^{+} }$          & $\mathrm{Ru   ^{++} }$         & $\mathrm{Rh   ^{} }$           & $\mathrm{Rh   ^{+} }$          & $\mathrm{Rh   ^{++} }$         \\
$\mathrm{Pd   ^{} }$           & $\mathrm{Pd   ^{+} }$          & $\mathrm{Pd   ^{++} }$         & $\mathrm{Ag   ^{} }$           & $\mathrm{Ag   ^{+} }$          & $\mathrm{Ag   ^{++} }$         & $\mathrm{Cd   ^{} }$           & $\mathrm{Cd   ^{+} }$          & $\mathrm{Cd   ^{++} }$         & $\mathrm{In   ^{} }$           & $\mathrm{In   ^{+} }$          & $\mathrm{In   ^{++} }$         & $\mathrm{Sn   ^{} }$           \\
$\mathrm{Sn   ^{+} }$          & $\mathrm{Sn   ^{++} }$         & $\mathrm{Sb   ^{} }$           & $\mathrm{Sb   ^{+} }$          & $\mathrm{Sb   ^{++} }$         & $\mathrm{Te   ^{} }$           & $\mathrm{Te   ^{+} }$          & $\mathrm{Te   ^{++} }$         & $\mathrm{I   ^{} }$            & $\mathrm{I   ^{+} }$           & $\mathrm{I   ^{++} }$          & $\mathrm{Xe   ^{} }$           & $\mathrm{Xe   ^{+} }$          \\
$\mathrm{Xe   ^{++} }$         & $\mathrm{Cs   ^{} }$           & $\mathrm{Cs   ^{+} }$          & $\mathrm{Cs   ^{++} }$         & $\mathrm{Ba   ^{} }$           & $\mathrm{Ba   ^{+} }$          & $\mathrm{Ba   ^{++} }$         & $\mathrm{La   ^{} }$           & $\mathrm{La   ^{+} }$          & $\mathrm{La   ^{++} }$         & $\mathrm{Ce   ^{} }$           & $\mathrm{Ce   ^{+} }$          & $\mathrm{Ce   ^{++} }$         \\
$\mathrm{Pr   ^{} }$           & $\mathrm{Pr   ^{+} }$          & $\mathrm{Pr   ^{++} }$         & $\mathrm{Nd   ^{} }$           & $\mathrm{Nd   ^{+} }$          & $\mathrm{Nd   ^{++} }$         & $\mathrm{Pm   ^{} }$           & $\mathrm{Pm   ^{+} }$          & $\mathrm{Pm   ^{++} }$         & $\mathrm{Sm   ^{} }$           & $\mathrm{Sm   ^{+} }$          & $\mathrm{Sm   ^{++} }$         & $\mathrm{Eu   ^{} }$           \\
$\mathrm{Eu   ^{+} }$          & $\mathrm{Eu   ^{++} }$         & $\mathrm{Gd   ^{} }$           & $\mathrm{Gd   ^{+} }$          & $\mathrm{Gd   ^{++} }$         & $\mathrm{Tb   ^{} }$           & $\mathrm{Tb   ^{+} }$          & $\mathrm{Tb   ^{++} }$         & $\mathrm{Dy   ^{} }$           & $\mathrm{Dy   ^{+} }$          & $\mathrm{Dy   ^{++} }$         & $\mathrm{Ho   ^{} }$           & $\mathrm{Ho   ^{+} }$          \\
$\mathrm{Ho   ^{++} }$         & $\mathrm{Er   ^{} }$           & $\mathrm{Er   ^{+} }$          & $\mathrm{Er   ^{++} }$         & $\mathrm{Tm   ^{} }$           & $\mathrm{Tm   ^{+} }$          & $\mathrm{Tm   ^{++} }$         & $\mathrm{Yb   ^{} }$           & $\mathrm{Yb   ^{+} }$          & $\mathrm{Yb   ^{++} }$         & $\mathrm{Lu   ^{} }$           & $\mathrm{Lu   ^{+} }$          & $\mathrm{Lu   ^{++} }$         \\
$\mathrm{Hf   ^{} }$           & $\mathrm{Hf   ^{+} }$          & $\mathrm{Hf   ^{++} }$         & $\mathrm{Ta   ^{} }$           & $\mathrm{Ta   ^{+} }$          & $\mathrm{Ta   ^{++} }$         & $\mathrm{W   ^{} }$            & $\mathrm{W   ^{+} }$           & $\mathrm{W   ^{++} }$          & $\mathrm{Re   ^{} }$           & $\mathrm{Re   ^{+} }$          & $\mathrm{Re   ^{++} }$         & $\mathrm{Os   ^{} }$           \\
$\mathrm{Os   ^{+} }$          & $\mathrm{Os   ^{++} }$         & $\mathrm{Ir   ^{} }$           & $\mathrm{Ir   ^{+} }$          & $\mathrm{Ir   ^{++} }$         & $\mathrm{Pt   ^{} }$           & $\mathrm{Pt   ^{+} }$          & $\mathrm{Pt   ^{++} }$         & $\mathrm{Au   ^{} }$           & $\mathrm{Au   ^{+} }$          & $\mathrm{Au   ^{++} }$         & $\mathrm{Hg   ^{} }$           & $\mathrm{Hg   ^{+} }$          \\
$\mathrm{Hg   ^{++} }$         & $\mathrm{Tl   ^{} }$           & $\mathrm{Tl   ^{+} }$          & $\mathrm{Tl   ^{++} }$         & $\mathrm{Pb   ^{} }$           & $\mathrm{Pb   ^{+} }$          & $\mathrm{Pb   ^{++} }$         & $\mathrm{Bi   ^{} }$           & $\mathrm{Bi   ^{+} }$          & $\mathrm{Bi   ^{++} }$         & $\mathrm{Po   ^{} }$           & $\mathrm{Po   ^{+} }$          & $\mathrm{Po   ^{++} }$         \\
$\mathrm{At   ^{} }$           & $\mathrm{At   ^{+} }$          & $\mathrm{At   ^{++} }$         & $\mathrm{Rn   ^{} }$           & $\mathrm{Rn   ^{+} }$          & $\mathrm{Rn   ^{++} }$         & $\mathrm{Fr   ^{} }$           & $\mathrm{Fr   ^{+} }$          & $\mathrm{Fr   ^{++} }$         & $\mathrm{Ra   ^{} }$           & $\mathrm{Ra   ^{+} }$          & $\mathrm{Ra   ^{++} }$         & $\mathrm{Ac   ^{} }$           \\
$\mathrm{Ac   ^{+} }$          & $\mathrm{Ac   ^{++} }$         & $\mathrm{Th   ^{} }$           & $\mathrm{Th   ^{+} }$          & $\mathrm{Th   ^{++} }$         & $\mathrm{Pa   ^{} }$           & $\mathrm{Pa   ^{+} }$          & $\mathrm{Pa   ^{++} }$         & $\mathrm{U   ^{} }$            & $\mathrm{U   ^{+} }$           & $\mathrm{U   ^{++} }$          & $\mathrm{H   ^{-} }$           & $\mathrm{H_2   ^{} }$          \\
$\mathrm{H_2   ^{+} }$         & $\mathrm{HF   ^{} }$           & $\mathrm{HF   ^{+} }$          & $\mathrm{HCl   ^{} }$          & $\mathrm{HCl   ^{+} }$         & $\mathrm{HBr   ^{} }$          & $\mathrm{HBr   ^{+} }$         & $\mathrm{HI   ^{} }$           & $\mathrm{HeH   ^{+} }$         & $\mathrm{He_2   ^{} }$         & $\mathrm{He_2   ^{+} }$        & $\mathrm{HeNe   ^{+} }$        & $\mathrm{LiH   ^{} }$          \\
$\mathrm{LiH   ^{+} }$         & $\mathrm{Li_2   ^{} }$         & $\mathrm{LiO   ^{} }$          & $\mathrm{LiF   ^{} }$          & $\mathrm{LiNa   ^{} }$         & $\mathrm{LiCl   ^{} }$         & $\mathrm{LiK   ^{} }$          & $\mathrm{LiBr   ^{} }$         & $\mathrm{LiI   ^{} }$          & $\mathrm{BeH   ^{} }$          & $\mathrm{BeH   ^{+} }$         & $\mathrm{BeO   ^{} }$          & $\mathrm{BeF   ^{} }$          \\
$\mathrm{BeS   ^{} }$          & $\mathrm{BeCl   ^{} }$         & $\mathrm{BH   ^{} }$           & $\mathrm{BH   ^{+} }$          & $\mathrm{B_2   ^{} }$          & $\mathrm{BO   ^{} }$           & $\mathrm{BF   ^{} }$           & $\mathrm{BS   ^{} }$           & $\mathrm{BCl   ^{} }$          & $\mathrm{BBr   ^{} }$          & $\mathrm{CH   ^{} }$           & $\mathrm{CH   ^{+} }$          & $\mathrm{C_2   ^{} }$          \\
$\mathrm{C_2   ^{+} }$         & $\mathrm{CN   ^{} }$           & $\mathrm{CN   ^{+} }$          & $\mathrm{CO   ^{} }$           & $\mathrm{CO   ^{+} }$          & $\mathrm{CF   ^{} }$           & $\mathrm{CP   ^{} }$           & $\mathrm{CS   ^{} }$           & $\mathrm{CS   ^{+} }$          & $\mathrm{CCl   ^{} }$          & $\mathrm{CSe   ^{} }$          & $\mathrm{CBr   ^{} }$          & $\mathrm{NH   ^{} }$           \\
$\mathrm{NH   ^{+} }$          & $\mathrm{N_2   ^{} }$          & $\mathrm{N_2   ^{+} }$         & $\mathrm{NO   ^{} }$           & $\mathrm{NO   ^{+} }$          & $\mathrm{NF   ^{} }$           & $\mathrm{NS   ^{} }$           & $\mathrm{NS   ^{+} }$          & $\mathrm{NSe   ^{} }$          & $\mathrm{NBr   ^{} }$          & $\mathrm{OH   ^{} }$           & $\mathrm{OH   ^{+} }$          & $\mathrm{O_2   ^{} }$          \\
$\mathrm{O_2   ^{+} }$         & $\mathrm{F   ^{-} }$           & $\mathrm{FO   ^{} }$           & $\mathrm{F_2   ^{} }$          & $\mathrm{F_2   ^{+} }$         & $\mathrm{NeH   ^{+} }$         & $\mathrm{Ne_2   ^{} }$         & $\mathrm{Ne_2   ^{+} }$        & $\mathrm{NaH   ^{} }$          & $\mathrm{NaO   ^{} }$          & $\mathrm{NaF   ^{} }$          & $\mathrm{Na_2   ^{} }$         & $\mathrm{Na_2   ^{+} }$        \\
$\mathrm{NaCl   ^{} }$         & $\mathrm{NaBr   ^{} }$         & $\mathrm{NaI   ^{} }$          & $\mathrm{MgH   ^{} }$          & $\mathrm{MgH   ^{+} }$         & $\mathrm{MgO   ^{} }$          & $\mathrm{MgF   ^{} }$          & $\mathrm{Mg_2   ^{} }$         & $\mathrm{MgS   ^{} }$          & $\mathrm{MgCl   ^{} }$         & $\mathrm{MgBr   ^{} }$         & $\mathrm{AlH   ^{} }$          & $\mathrm{AlO   ^{} }$          \\
$\mathrm{AlF   ^{} }$          & $\mathrm{Al_2   ^{} }$         & $\mathrm{AlS   ^{} }$          & $\mathrm{AlCl   ^{} }$         & $\mathrm{AlSe   ^{} }$         & $\mathrm{AlBr   ^{} }$         & $\mathrm{AlI   ^{} }$          & $\mathrm{SiH   ^{} }$          & $\mathrm{SiH   ^{+} }$         & $\mathrm{SiC   ^{} }$          & $\mathrm{SiO   ^{} }$          & $\mathrm{SiO   ^{+} }$         & $\mathrm{SiF   ^{} }$          \\
$\mathrm{Si_2   ^{} }$         & $\mathrm{SiS   ^{} }$          & $\mathrm{SiSe   ^{} }$         & $\mathrm{SiTe   ^{} }$         & $\mathrm{SiI   ^{} }$          & $\mathrm{PH   ^{} }$           & $\mathrm{PH   ^{+} }$          & $\mathrm{PN   ^{} }$           & $\mathrm{PO   ^{} }$           & $\mathrm{PO   ^{+} }$          & $\mathrm{P_2   ^{} }$          & $\mathrm{P_2   ^{+} }$         & $\mathrm{PS   ^{} }$           \\
$\mathrm{SH   ^{} }$           & $\mathrm{SH   ^{+} }$          & $\mathrm{SO   ^{} }$           & $\mathrm{SO   ^{+} }$          & $\mathrm{SF   ^{} }$           & $\mathrm{S_2   ^{} }$          & $\mathrm{S_2   ^{+} }$         & $\mathrm{Cl   ^{-} }$          & $\mathrm{ClO   ^{} }$          & $\mathrm{ClF   ^{} }$          & $\mathrm{Cl_2   ^{} }$         & $\mathrm{Cl_2   ^{+} }$        & $\mathrm{Ar_2   ^{} }$         \\
$\mathrm{KH   ^{} }$           & $\mathrm{KO   ^{} }$           & $\mathrm{KF   ^{} }$           & $\mathrm{KCl   ^{} }$          & $\mathrm{K_2   ^{} }$          & $\mathrm{KBr   ^{} }$          & $\mathrm{KI   ^{} }$           & $\mathrm{CaH   ^{} }$          & $\mathrm{CaO   ^{} }$          & $\mathrm{CaF   ^{} }$          & $\mathrm{CaS   ^{} }$          & $\mathrm{CaCl   ^{} }$         & $\mathrm{CaCa   ^{} }$         \\
$\mathrm{ScH   ^{} }$          & $\mathrm{ScO   ^{} }$          & $\mathrm{ScF   ^{} }$          & $\mathrm{ScS   ^{} }$          & $\mathrm{ScCl   ^{} }$         & $\mathrm{TiH   ^{} }$          & $\mathrm{TiN   ^{} }$          & $\mathrm{TiO   ^{} }$          & $\mathrm{TiS   ^{} }$          & $\mathrm{VO   ^{} }$           & $\mathrm{CrH   ^{} }$          & $\mathrm{CrO   ^{} }$          & $\mathrm{CrS   ^{} }$          \\
$\mathrm{MnH   ^{} }$          & $\mathrm{MnO   ^{} }$          & $\mathrm{MnF   ^{} }$          & $\mathrm{MnS   ^{} }$          & $\mathrm{MnCl   ^{} }$         & $\mathrm{FeH   ^{} }$          & $\mathrm{FeO   ^{} }$          & $\mathrm{CoH   ^{} }$          & $\mathrm{CoCl   ^{} }$         & $\mathrm{NiH   ^{} }$          & $\mathrm{NiO   ^{} }$          & $\mathrm{NiCl   ^{} }$         & $\mathrm{NiBr   ^{} }$         \\
$\mathrm{CuH   ^{} }$          & $\mathrm{CuO   ^{} }$          & $\mathrm{CuF   ^{} }$          & $\mathrm{CuS   ^{} }$          & $\mathrm{CuCl   ^{} }$         & $\mathrm{Cu_2   ^{} }$         & $\mathrm{CuSe   ^{} }$         & $\mathrm{CuBr   ^{} }$         & $\mathrm{CuTe   ^{} }$         & $\mathrm{CuI   ^{} }$          & $\mathrm{ZnH   ^{} }$          & $\mathrm{ZnH   ^{+} }$         & $\mathrm{ZnCl   ^{} }$         \\
$\mathrm{GaH   ^{} }$          & $\mathrm{GaO   ^{} }$          & $\mathrm{GaF   ^{} }$          & $\mathrm{GaCl   ^{} }$         & $\mathrm{GaBr   ^{} }$         & $\mathrm{GaI   ^{} }$          & $\mathrm{GeH   ^{} }$          & $\mathrm{GeO   ^{} }$          & $\mathrm{GeF   ^{} }$          & $\mathrm{GeS   ^{} }$          & $\mathrm{GeCl   ^{} }$         & $\mathrm{GeSe   ^{} }$         & $\mathrm{GeTe   ^{} }$         \\
$\mathrm{AsH   ^{} }$          & $\mathrm{AsO   ^{} }$          & $\mathrm{AsF   ^{} }$          & $\mathrm{AsAs   ^{} }$         & $\mathrm{SeH   ^{} }$          & $\mathrm{SeO   ^{} }$          & $\mathrm{SeF   ^{} }$          & $\mathrm{SeS   ^{} }$          & $\mathrm{Se_2   ^{} }$         & $\mathrm{BrO   ^{} }$          & $\mathrm{BrF   ^{} }$          & $\mathrm{BrCl   ^{} }$         & $\mathrm{Br_2   ^{} }$         \\
$\mathrm{Br_2   ^{+} }$        & $\mathrm{KrF   ^{+} }$         & $\mathrm{RbF   ^{} }$          & $\mathrm{RbCl   ^{} }$         & $\mathrm{RbBr   ^{} }$         & $\mathrm{RbI   ^{} }$          & $\mathrm{SrH   ^{} }$          & $\mathrm{SrO   ^{} }$          & $\mathrm{SrF   ^{} }$          & $\mathrm{SrS   ^{} }$          & $\mathrm{SrCl   ^{} }$         & $\mathrm{YO   ^{} }$           & $\mathrm{YF   ^{} }$           \\
$\mathrm{YS   ^{} }$           & $\mathrm{ZrN   ^{} }$          & $\mathrm{ZrO   ^{} }$          & $\mathrm{NbO   ^{} }$          & $\mathrm{RuC   ^{} }$          & $\mathrm{RuO   ^{} }$          & $\mathrm{RhC   ^{} }$          & $\mathrm{AgH   ^{} }$          & $\mathrm{AgO   ^{} }$          & $\mathrm{AgF   ^{} }$          & $\mathrm{AgAl   ^{} }$         & $\mathrm{AgCl   ^{} }$         & $\mathrm{AgBr   ^{} }$         \\
$\mathrm{AgI   ^{} }$          & $\mathrm{CdH   ^{} }$          & $\mathrm{CdH   ^{+} }$         & $\mathrm{CdF   ^{} }$          & $\mathrm{CdCl   ^{} }$         & $\mathrm{InH   ^{} }$          & $\mathrm{InO   ^{} }$          & $\mathrm{InF   ^{} }$          & $\mathrm{InCl   ^{} }$         & $\mathrm{InBr   ^{} }$         & $\mathrm{InI   ^{} }$          & $\mathrm{SnH   ^{} }$          & $\mathrm{SnO   ^{} }$          \\
$\mathrm{SnF   ^{} }$          & $\mathrm{SnS   ^{} }$          & $\mathrm{SnSe   ^{} }$         & $\mathrm{SnTe   ^{} }$         & $\mathrm{SbO   ^{} }$          & $\mathrm{SbF   ^{} }$          & $\mathrm{SbP   ^{} }$          & $\mathrm{SbSb   ^{} }$         & $\mathrm{TeO   ^{} }$          & $\mathrm{TeS   ^{} }$          & $\mathrm{TeSe   ^{} }$         & $\mathrm{Te_2   ^{} }$         & $\mathrm{IO   ^{} }$           \\
$\mathrm{IF   ^{} }$           & $\mathrm{ICl   ^{} }$          & $\mathrm{IBr   ^{} }$          & $\mathrm{II   ^{} }$           & $\mathrm{Xe_2   ^{} }$         & $\mathrm{CsH   ^{} }$          & $\mathrm{CsF   ^{} }$          & $\mathrm{CsCl   ^{} }$         & $\mathrm{CsBr   ^{} }$         & $\mathrm{CsI   ^{} }$          & $\mathrm{Cs_2   ^{} }$         & $\mathrm{BaH   ^{} }$          & $\mathrm{BaO   ^{} }$          \\
$\mathrm{BaF   ^{} }$          & $\mathrm{BaS   ^{} }$          & $\mathrm{BaCl   ^{} }$         & $\mathrm{LaO   ^{} }$          & $\mathrm{LaS   ^{} }$          & $\mathrm{CeO   ^{} }$          & $\mathrm{PrO   ^{} }$          & $\mathrm{TbO   ^{} }$          & $\mathrm{HoF   ^{} }$          & $\mathrm{YbH   ^{} }$          & $\mathrm{YbF   ^{} }$          & $\mathrm{LuO   ^{} }$          & $\mathrm{LuF   ^{} }$          \\
$\mathrm{HfO   ^{} }$          & $\mathrm{TaO   ^{} }$          & $\mathrm{TaO   ^{+} }$         & $\mathrm{WO   ^{} }$           & $\mathrm{IrC   ^{} }$          & $\mathrm{IrO   ^{} }$          & $\mathrm{PtH   ^{} }$          & $\mathrm{PtC   ^{} }$          & $\mathrm{PtO   ^{} }$          & $\mathrm{AuH   ^{} }$          & $\mathrm{Au_2   ^{} }$         & $\mathrm{AuSi   ^{} }$         & $\mathrm{AuCl   ^{} }$         \\
$\mathrm{Au_2   ^{} }$         & $\mathrm{HgH   ^{} }$          & $\mathrm{HgH   ^{+} }$         & $\mathrm{HgF   ^{} }$          & $\mathrm{HgCl   ^{} }$         & $\mathrm{HgAr   ^{+} }$        & $\mathrm{TlH   ^{} }$          & $\mathrm{TlF   ^{} }$          & $\mathrm{TlCl   ^{} }$         & $\mathrm{TlBr   ^{} }$         & $\mathrm{TlI   ^{} }$          & $\mathrm{PbH   ^{} }$          & $\mathrm{PbO   ^{} }$          \\
$\mathrm{PbF   ^{} }$          & $\mathrm{PbS   ^{} }$          & $\mathrm{PbCl   ^{} }$         & $\mathrm{PbSe   ^{} }$         & $\mathrm{PbTe   ^{} }$         & $\mathrm{BiH   ^{} }$          & $\mathrm{BiO   ^{} }$          & $\mathrm{BiS   ^{} }$          & $\mathrm{BiCl   ^{} }$         & $\mathrm{Bi_2   ^{} }$         & $\mathrm{ThO   ^{} }$          & $\mathrm{H_2   ^{-} }$         & $\mathrm{H_3   ^{+} }$         \\
$\mathrm{H_2O   ^{} }$         & $\mathrm{H_2S   ^{} }$         & $\mathrm{HBO   ^{} }$          & $\mathrm{HBS   ^{} }$          & $\mathrm{HCN   ^{} }$          & $\mathrm{HCO   ^{} }$          & $\mathrm{HCO   ^{+} }$         & $\mathrm{LiOH   ^{} }$         & $\mathrm{BeH_2   ^{} }$        & $\mathrm{Be_2O   ^{} }$        & $\mathrm{BeC_2   ^{} }$        & $\mathrm{BeOH   ^{} }$         & $\mathrm{BeF_2   ^{} }$        \\
$\mathrm{BeClF   ^{} }$        & $\mathrm{BeCl_2   ^{} }$       & $\mathrm{BH_2   ^{} }$         & $\mathrm{BO_2   ^{} }$         & $\mathrm{CH_2   ^{} }$         & $\mathrm{CHF   ^{} }$          & $\mathrm{CHP   ^{} }$          & $\mathrm{CHCl   ^{} }$         & $\mathrm{C_2   ^{-} }$         & $\mathrm{C_2H   ^{} }$         & $\mathrm{C_3   ^{} }$          & $\mathrm{C_2O   ^{} }$         & $\mathrm{CN   ^{-} }$          \\
$\mathrm{CNC   ^{} }$          & $\mathrm{CO_2   ^{} }$         & $\mathrm{COS   ^{} }$          & $\mathrm{CS_2   ^{} }$         & $\mathrm{NH_2   ^{} }$         & $\mathrm{NO   ^{-} }$          & $\mathrm{OH   ^{-} }$          & $\mathrm{OBF   ^{} }$          & $\mathrm{O_3   ^{} }$          & $\mathrm{OAlH   ^{} }$         & $\mathrm{OAlF   ^{} }$         & $\mathrm{OAlCl   ^{} }$        & $\mathrm{OTiF   ^{} }$         \\
$\mathrm{OTiCl   ^{} }$        & $\mathrm{F_2   ^{-} }$         & $\mathrm{NaCN   ^{} }$         & $\mathrm{NaOH   ^{} }$         & $\mathrm{MgOH   ^{} }$         & $\mathrm{MgF_2   ^{} }$        & $\mathrm{MgClF   ^{} }$        & $\mathrm{MgCl_2   ^{} }$       & $\mathrm{MgBr_2   ^{} }$       & $\mathrm{AlOH   ^{} }$         & $\mathrm{AlOH   ^{+} }$        & $\mathrm{AlF_2   ^{} }$        & $\mathrm{Al_2O   ^{} }$        \\
$\mathrm{AlClF   ^{} }$        & $\mathrm{AlCl_2   ^{} }$       & $\mathrm{SiH   ^{} }$          & $\mathrm{SiC_2   ^{} }$        & $\mathrm{SiO_2   ^{} }$        & $\mathrm{Si_2C   ^{} }$        & $\mathrm{Si_2N   ^{} }$        & $\mathrm{Si_3   ^{} }$         & $\mathrm{PH_2   ^{} }$         & $\mathrm{PO_2   ^{} }$         & $\mathrm{SH   ^{-} }$          & $\mathrm{SO_2   ^{} }$         & $\mathrm{ClCN   ^{} }$         \\
$\mathrm{KCN   ^{} }$          & $\mathrm{KOH   ^{} }$          & $\mathrm{CaOH   ^{} }$         & $\mathrm{CaOH   ^{+} }$        & $\mathrm{CaF_2   ^{} }$        & $\mathrm{CaCl_2   ^{} }$       & $\mathrm{CaBr_2   ^{} }$       & $\mathrm{CaI_2   ^{} }$        & $\mathrm{TiO_2   ^{} }$        & $\mathrm{TiF_2   ^{} }$        & $\mathrm{TiCl_2   ^{} }$       & $\mathrm{TiBr_2   ^{} }$       & $\mathrm{TiI_2   ^{} }$        \\
$\mathrm{VO_2   ^{} }$         & $\mathrm{CrO_2   ^{} }$        & $\mathrm{FeCl_2   ^{} }$       & $\mathrm{FeBr_2   ^{} }$       & $\mathrm{BrCN   ^{} }$         & $\mathrm{SrOH   ^{} }$         & $\mathrm{SrOH   ^{+} }$        & $\mathrm{SrF_2   ^{} }$        & $\mathrm{SrCl_2   ^{} }$       & $\mathrm{SrBr_2   ^{} }$       & $\mathrm{ZrO_2   ^{} }$        & $\mathrm{ZrF_2   ^{} }$        & $\mathrm{ZrCl_2   ^{} }$       \\
$\mathrm{ZrBr_2   ^{} }$       & $\mathrm{ZrI_2   ^{} }$        & $\mathrm{NbO_2   ^{} }$        & $\mathrm{MoO_2   ^{} }$        & $\mathrm{CsOH   ^{} }$         & $\mathrm{BaOH   ^{} }$         & $\mathrm{BaOH   ^{+} }$        & $\mathrm{BaF_2   ^{} }$        & $\mathrm{BaCl_2   ^{} }$       & $\mathrm{BaBr_2   ^{} }$       & $\mathrm{TaO_2   ^{} }$        & $\mathrm{WO_2   ^{} }$         & $\mathrm{WCl_2   ^{} }$        \\
$\mathrm{H_3O   ^{+} }$        & $\mathrm{HBO_2   ^{} }$        & $\mathrm{LiBO_2   ^{} }$       & $\mathrm{BeBO_2   ^{} }$       & $\mathrm{BH_3   ^{} }$         & $\mathrm{BO_2   ^{-} }$        & $\mathrm{CH_3   ^{} }$         & $\mathrm{C_2H_2   ^{} }$       & $\mathrm{C_2HF   ^{} }$        & $\mathrm{C_2HCl   ^{} }$       & $\mathrm{C_4   ^{} }$          & $\mathrm{(CN)_2   ^{} }$       & $\mathrm{CO_2   ^{-} }$        \\
$\mathrm{NH_3   ^{} }$         & $\mathrm{OAlOH   ^{} }$        & $\mathrm{OAlF_2   ^{} }$       & $\mathrm{NaBO_2   ^{} }$       & $\mathrm{(NaCl)_2   ^{} }$     & $\mathrm{(NaBr)_2   ^{} }$     & $\mathrm{AlBO_2   ^{} }$       & $\mathrm{AlOH   ^{-} }$        & $\mathrm{AlO_2   ^{-} }$       & $\mathrm{(AlO)_2   ^{} }$      & $\mathrm{AlF_2   ^{-} }$       & $\mathrm{AlCl_2   ^{-} }$      & $\mathrm{PH_3   ^{} }$         \\
$\mathrm{P_4   ^{} }$          & $\mathrm{KBO_2   ^{} }$        & $\mathrm{(KCl)_2   ^{} }$      & $\mathrm{(KBr)_2   ^{} }$      & $\mathrm{TiOCl_2   ^{} }$      & $\mathrm{TiF_3   ^{} }$        & $\mathrm{TiCl_3   ^{} }$       & $\mathrm{ZrCl_3   ^{} }$       & $\mathrm{MoO_3   ^{} }$        & $\mathrm{WO_3   ^{} }$         & $\mathrm{Be(OH)_2   ^{} }$     & $\mathrm{B(OH)_2   ^{} }$      & $\mathrm{CH_4   ^{} }$         \\
$\mathrm{CH_3Cl   ^{} }$       & $\mathrm{C_5   ^{} }$          & $\mathrm{OAlF_2   ^{-} }$      & $\mathrm{Mg(OH)_2   ^{} }$     & $\mathrm{SiH_4   ^{} }$        & $\mathrm{SiH_3F   ^{} }$       & $\mathrm{SiH_3Cl   ^{} }$      & $\mathrm{SiH_3Br   ^{} }$      & $\mathrm{SiH_3I   ^{} }$       & $\mathrm{SiH_2F_2   ^{} }$     & $\mathrm{Ca(OH)_2   ^{} }$     & $\mathrm{Fe(OH)_2   ^{} }$     & $\mathrm{Sr(OH)_2   ^{} }$     \\
$\mathrm{ZrF_4   ^{} }$        & $\mathrm{ZrCl_4   ^{} }$       & $\mathrm{Ba(OH)_2   ^{} }$     & $\mathrm{WO_2Cl_2   ^{} }$     & $\mathrm{(LiOH)_2   ^{} }$     & $\mathrm{(BeO)_3   ^{} }$      & $\mathrm{C_2H_4   ^{} }$       & $\mathrm{(NaCN)_2   ^{} }$     & $\mathrm{(NaOH)_2   ^{} }$     & $\mathrm{AlF_4   ^{-} }$       & $\mathrm{(KOH)_2   ^{} }$      & $\mathrm{H_3BO_3   ^{} }$      & $\mathrm{H_2MoO_4   ^{} }$     \\
$\mathrm{O_2W(OH)_2   ^{} }$   &                                &                                &                                &                                &                                &                                &                                &                                &                                &                                &                                &                                \\
\hline
\end{tabular}
}
\label{tab:polyatomic}
\end{table}

\end{document}